\documentclass[10pt,letterpaper,final,twocolumn,journal]{IEEEtran}

\usepackage{setspace}
\usepackage{nomencl}
\makenomenclature
\usepackage{graphicx}

\usepackage{soul}

\usepackage{algorithm}
\usepackage{algorithmic}
\usepackage{epstopdf}
\usepackage{epsfig}
\usepackage[cmex10]{amsmath}
\usepackage{cite}
\usepackage{amssymb}
\usepackage[usenames,dvipsnames]{color} 
\usepackage{multirow}
\usepackage{array}
\long\def\symbolfootnote[#1]#2{\begingroup\def\thefootnote{\fnsymbol{footnote}}\footnote[#1]{#2}\endgroup}
\usepackage[table]{xcolor}
\usepackage{tabularx}
\usepackage{chngcntr}
\usepackage{booktabs}
\usepackage{caption}
\usepackage{subcaption}
\usepackage{float}
\usepackage{mathtools,cuted}
\usepackage{braket}

\usepackage{hyperref}

\usepackage{amsthm}
\theoremstyle{definition}
\newtheorem{definition}{Definition}
\newtheorem{proposition}{Proposition}

\newtheorem*{remark}{Remark}

\newcommand{\subparagraph}{}

\usepackage{braket}

\begin{document}

\markboth{IEEE Transactions on Communications}{Accepted paper}

\title{Direct Quantum Communications in the Presence of Realistic Noisy Entanglement}
\author{Daryus Chandra,~\IEEEmembership{Member,~IEEE}, Angela~Sara~Cacciapuoti,~\IEEEmembership{Senior Member,~IEEE}, Marcello~Caleffi,~\IEEEmembership{Senior Member,~IEEE}, Lajos~Hanzo,~\IEEEmembership{Fellow,~IEEE}
   	\thanks{D. Chandra and L. Hanzo are with the School of Electronics and Computer Science (ECS), University of Southampton, Southampton, SO17 1BJ, UK. Email: \href{mailto:dc3c18@ecs.soton.ac.uk}{dc3c18@ecs.soton.ac.uk}, \href{mailto:lh@ecs.soton.ac.uk}{lh@ecs.soton.ac.uk}. Web: \href{www-mobile.ecs.soton.ac.uk}{www-mobile.ecs.soton.ac.uk}.}
	\thanks{M. Caleffi and A.S. Cacciapuoti are with the Department of Electrical Engineering and Information Technology, University of Naples Federico II, Naples, 80125, Italy. E-mail: \href{mailto:marcello.caleffi@unina.it}{marcello.caleffi@unina.it}, \href{mailto:angelasara.cacciapuoti@unina.it}{angelasara.cacciapuoti@unina.it}. Web: \href{http://www.quantuminternet.it}{www.quantuminternet.it}.}
    \thanks{A.S. Cacciapuoti and M. Caleffi are also with the Laboratorio Nazionale di Comunicazioni Multimediali, National Inter-University Consortium for Telecommunications (CNIT), Naples, 80126, Italy.}
    \thanks{A.S Cacciapuoti, M. Caleffi and D. Chandra would like to acknowledge the financial support of the project ``Towards the Quantum Internet: A Multidisciplinary Effort'', University of Naples Federico II. A.S Cacciapuoti and M. Caleffi would like to to acknowledge as well the financial support of PON project ``S4E - Sistemi di Sicurezza e Protezione per l’Ambiente Mare''.}
	\thanks{L. Hanzo would like to acknowledge the financial support of the Engineering and Physical Sciences Research Council projects EP/P034284/1 and EP/P003990/1 (COALESCE) as well as of the European Research Council's Advanced Fellow Grant QuantCom (Grant No. 789028)}
}

\maketitle


\begin{abstract} 
To realize the Quantum Internet, quantum communications require pre-shared entanglement among quantum nodes. However, both the generation and the distribution of the maximally-entangled quantum states are inherently contaminated by quantum decoherence. Conventionally, the quantum decoherence is mitigated by performing the consecutive steps of quantum entanglement distillation followed by quantum teleportation. However, this conventional approach imposes a long delay. To circumvent this impediment, we propose a novel quantum communication scheme relying on realistic noisy pre-shared entanglement, which eliminates the sequential steps imposing delay in the standard approach. More precisely, our proposed scheme can be viewed as a direct quantum communication scheme capable of improving the quantum bit error ratio (QBER) of the logical qubits despite relying on realistic noisy pre-shared entanglement. Our performance analysis shows that the proposed scheme offers competitive QBER, yield, and goodput compared to the existing state-of-the-art quantum communication schemes, despite requiring fewer quantum gates.
\end{abstract}


\pagenumbering{arabic}


\section{Introduction}
\label{Introduction}

Enabling quantum communications among quantum devices within the Quantum Internet~\cite{kimble2008quantum,caleffi2018quantum,wehner2018quantum} will ultimately lead to various groundbreaking applications. These radically new applications do not necessarily have classical counterparts~\cite{qirg-use} and they are not limited to the already well-known secure classical communications, blind  computation, distributed quantum computing, and quantum secret sharing~\cite{razavi2012multiple, caleffi2018quantum, cuomo2020towards, caleffi2020rise, sun2020toward}. Naturally, the reliable transfer of quantum information is sought across the quantum network relying on quantum channels~\cite{cacciapuoti2019quantum, cacciapuoti2020entanglement}. However, the quantum channels inevitably impose deleterious quantum decoherence, which inflicts quantum errors~\cite{nielsen2000quantum, cariolaro2010performance}. In the classical domain, the errors imposed by the communication channels can be mitigated using error-control codes~\cite{shannon1948mathematical}. The key idea of error-control codes is to attach appropriately designed redundancy to the information bits by an encoding process, which is utilized by the decoder to correct a certain number of errors. However, observing and/or copying quantum information is not allowed in the quantum domain due to the no-cloning theorem and the quantum measurement postulate. This motivates the carefully constructed design of quantum error-correction codes (QECCs)~\cite{shor1995scheme, calderbank1996good, laflamme1996perfect, lidar2013quantum}.

QECCs constitute potent error mitigation techniques required for tackling the deleterious effect of quantum decoherence. Similar to the classical error-correction codes, QECCs rely on attaching redundant qubits to the logical qubits to provide additional information that can be exploited for quantum error-correction during the decoding step~\cite{babar2018duality}. Interestingly, the whole encoding and decoding process can be completed without actually observing the physical qubits and thus, preserving the integrity of the quantum information conveyed by the physical qubits. In the quantum domain, the redundant qubits can be in form of auxiliary qubits initialized to the $\ket{0}$ or $\ket{+}$ states, or in the form of pre-shared maximally-entangled quantum states, which are normally assumed to be noise-free. For a two-qubit system, the maximally-entangled quantum states are represented by the Einstein-Podolsky-Rosen (EPR) pairs. The state-of-the-art studies typically assume that the EPR pairs are pre-shared among quantum devices within the quantum networks before any quantum communication protocol is initiated. Hence, the EPR pairs can be considered as the primary resource within the Quantum Internet~\cite{cacciapuoti2020entanglement}.

Having pre-shared entanglement offers several beneficial features for QECCs. Firstly, it can be used for conveniently transforming some powerful classical error-correction codes that do not satisfy the symplectic criterion\footnote{A pair of classical error-correction codes having parity-check matrices $H_x$ and $H_z$ can be transformed to a quantum error-correction code if they satisfy $H_x H_z^T + H_z H_x^T  = 0 \mod 2$.} into their quantum counterparts~\cite{fujiwara2010entanglement, wilde2012quantum, wilde2013entanglement}. Secondly, they can also be used for increasing the error-correction capability of quantum stabilizer codes (QSCs)~\cite{chandra2017quantum}. Indeed, there are several types of QECCs in the literature that exploit pre-shared entanglement, such as entanglement-assisted QSCs~\cite{brun2006correcting}, entanglement-aided canonical codes~\cite{devetak2009entanglement}, as well as teleportation-based QECCs~\cite{grassl2016entanglement}. However, in all the above-mentioned schemes, the pre-shared entanglement is considered to be noise-free.

In a scenario having realistic noisy pre-shared entanglement, QECCs are invoked for quantum entanglement distillation (QED)~\cite{bennett1996concentrating, bennett1996purification, bennett1996mixed, matsumoto2003conversion, lidar2013quantum}, which is followed by quantum teleportation~\cite{bennett1993teleporting} for transferring the quantum information. QED can be viewed as a specific application of QECCs, where several copies of noisy pre-shared EPR pairs are discarded to obtain fewer but less noisy EPR pairs. In this approach, QED and quantum teleportation have to be performed subsequently, which typically imposes excessive practical delay. Additionally, state-of-the-art QED schemes will always have some residual quantum noise, unless infinitely many noisy pre-shared EPR pairs are discarded during QED. Unfortunately, this residual quantum noise is carried over to the logical qubits during the quantum teleportation process and hence it affects the integrity of the quantum information. In this treatise, we refer to this specific quantum communication scheme relying on the consecutive steps of QED and quantum teleportation as QED+QT.

Another QECC-aided technique operating in the presence of noisy pre-shared entanglement was introduced in~\cite{lai2012entanglement}, which we will refer to as quantum stabilizer codes using imperfect pre-shared entanglement (QSC-IE). Compared to the QED-based schemes which apply the QECCs locally on the pre-shared EPR pairs split between the transmitter and the receiver, the scheme presented in~\cite{lai2012entanglement} requires that the pre-shared portion of the EPR pairs at the transmitter side has to be sent to the receiver in order to apply stabilizer measurements to both qubits of the EPR pairs. Consequently, a relatively high number of the two-qubit quantum gates -- as exemplified by quantum controlled-NOT (CNOT) gate -- are required for performing these measurements. Furthermore, by reasoning for a fixed number of pre-shared EPR pairs and logical qubits, the QSC-IE scheme demands for a higher number of quantum channel uses, to apply stabilizer measurements to both qubits of the EPR pairs.

Having said that, in this treatise, we propose a novel solution for achieving a reliable quantum communication, despite using noisy pre-shared entanglement. Firstly, we eliminate the idealized simplifying assumption of having noise-free pre-shared EPR pairs. Secondly, we devise a scheme for avoiding the undesired delay imposed by the consecutive steps of QED and quantum teleportation in conventional twin-step QED+QT schemes. By contrast, our proposed scheme can be viewed as a single-step direct quantum communication scheme, which exploits the quantum noise experienced by the pre-shared EPR pairs for improving the reliability of quantum communications by encoding the logical qubits directly with the aid of noisy pre-shared EPR pairs. As it will become more evident later in this treatise, our proposal may be deemed philosophically reminiscent of training-based equalization techniques in classical communications, which rely on pilot sequences for estimating the channel and then eliminating its impairments. Thirdly, we also eliminate the necessity of performing stabilizer measurements on both qubits of the pre-shared EPR pairs for the sake of reducing: (i) the number of quantum gates required to achieve reliable quantum communications and (ii) the uses of the considered quantum channel. Indeed, by relying solely on the local measurements of the pre-shared EPR pairs, our proposal significantly reduces the number of the required two-qubit quantum gates as well as the number of quantum channel uses by reasoning with the same number of pre-shared EPR pairs. Table~\ref{table:proposed} boldly and explicitly contrasts our proposed scheme to the existing schemes of amalgamating pre-shared entanglement and QECCs. Naturally, our proposal is also suitable for the scenario of noise-free pre-shared entanglement, similarly to the EA-QECC schemes. In Section~\ref{Discussion: A Quantum Computing Perspective}, we formally show that our proposed scheme outperforms the state-of-the-art. 

\begin{table*}
	\caption{Comparison of our proposed scheme with the state-of-the-art schemes.}
	\small
	\centering
	\begin{tabular}{|c|c|c|c|}
	\hline
	Scheme & Noisy Pre-Shared Entanglement & Direct Communication & Stabilizer Measurement \\
	\hline
	\hline
	QED+QT & Yes & No & Yes \\
	\hline
	QSC-IE & Yes & Yes & Yes \\
	\hline
	EA-QECC & No & Yes & Yes \\
	\hline
	{\bf Proposed} & {\bf Yes} & {\bf Yes} & {\bf No} \\
	\hline
	\end{tabular}
	\label{table:proposed}
\end{table*}

Our novel contributions can be summarized as follows:
\begin{enumerate}
    \item We propose a new scheme for achieving reliable quantum communications despite relying on noisy pre-shared entanglement. More specifically, 
    \item We carry out the performance analysis of the proposed scheme for both error-detection and error-correction based schemes over quantum depolarizing channels. The results show that the proposed scheme offers competitive performance in terms of its qubit error ratio, yield, and goodput despite requiring fewer quantum gates than the existing state-of-the-art schemes.
    \item In case of noise-free pre-shared entanglement, the proposed scheme outperforms even the existing entanglement-assisted quantum stabilizer codes.
\end{enumerate}

The rest of the treatise is organized as follows. In Section~\ref{System Model}, we commence by presenting the quantum communication model. In Section~\ref{Quantum Communication with Noisy Pre-Shared Entanglement}, we detail the explicit formulation of our proposed scheme for direct noiseless quantum communication over noisy pre-shared entanglement. In Section~\ref{Error-Detection Scheme}, we exemplify our scheme proposed for error-detection, while in Section~\ref{Error-Correction Scheme}, we conceive its counterpart for error-correction. In Section~\ref{Discussion: A Quantum Computing Perspective}, we show the suitability of our proposal for quantum computing applications. Finally, we conclude in Section~\ref{Conclusions and Future Works} by also discussing some future research directions.


\section{System Model}
\label{System Model}

\begin{figure}[t]
    \center
    \includegraphics[width=\columnwidth]{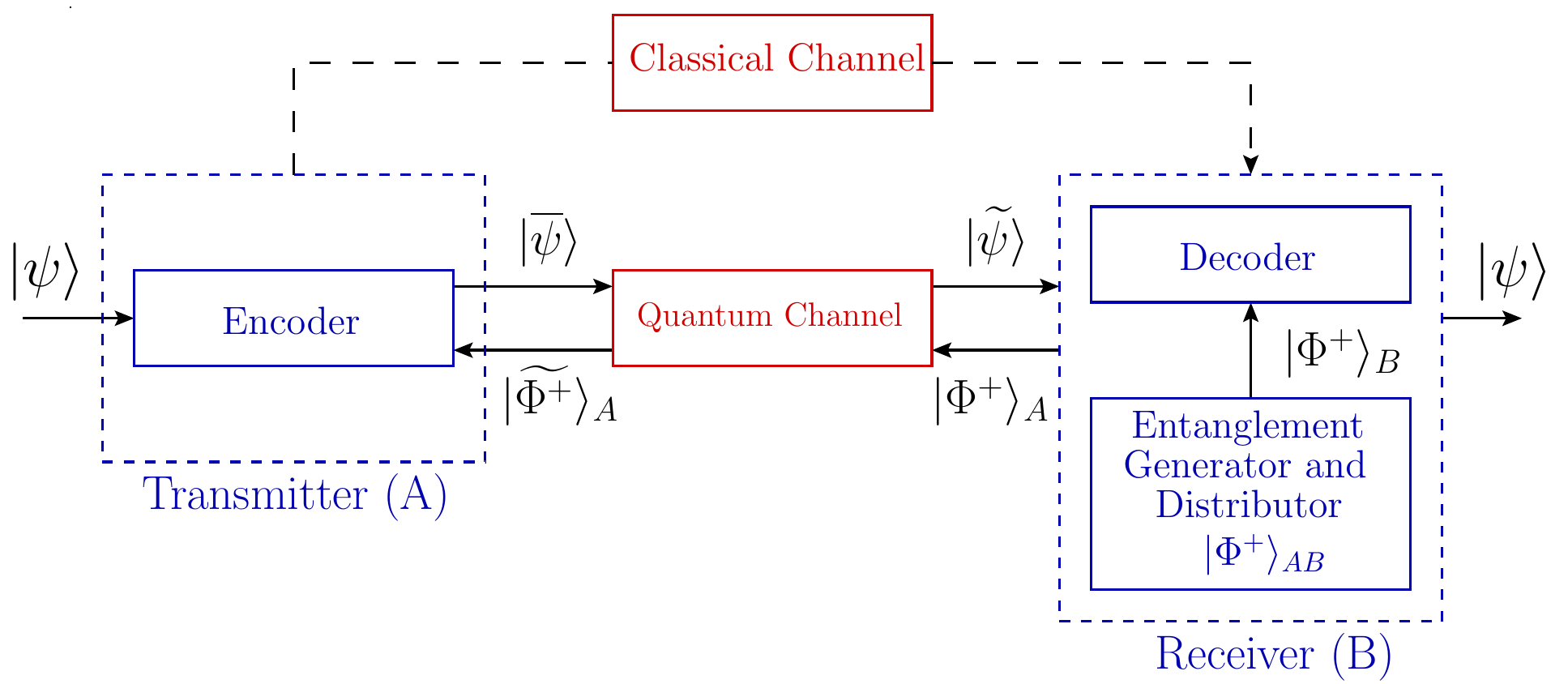}
	\caption{The quantum communication model considered for our proposed scheme.}
	\label{fig:model}
\end{figure}

As discussed in~\cite{cacciapuoti2020entanglement}, both entanglement generation and distribution are the key for the Quantum Internet. The specific ``location'' of the device implementing these functionalities -- a.k.a. the \textit{entanglement generator and distributor} -- varies among the different schemes and solutions~\cite{cacciapuoti2020entanglement}. However, there is a general agreement in the literature that the employment of the so-called ``\textit{at both end-points}'' scheme is vital for the Quantum Internet by enabling on-demand communication capabilities at the quantum nodes. According to the ``\textit{at both end-points}'' scheme, the entanglement generator and distributor is embedded within both the transmitter and the receiver~\cite{cacciapuoti2020entanglement}. In this light, we consider the quantum communication model depicted in Fig.~\ref{fig:model}. The model includes a transmitter $(A)$, a receiver $(B)$, the entanglement generator and distributor, a noisy quantum channel and a classical channel. Without loss in generality, in the figure we only highlight the entanglement generator and distributor used at the receiver, since it is exploited by the proposed scheme. The quantum communication session commences with the generation of the EPR pairs, whose quantum state is
\begin{align}
	\ket{\Phi^{\pm}}_{AB} &= \frac{1}{\sqrt{2}}\left(\ket{00} \pm \ket{11}_{AB}\right), \nonumber \\
	\ket{\Psi^{\pm}}_{AB} &= \frac{1}{\sqrt{2}}\left(\ket{01} \pm \ket{10}_{AB}\right).
\end{align}
In the rest of this treatise, we assume that the pre-shared EPR pairs are initialized to the quantum state of $\ket{\Phi^+}_{AB}$, where the subscript $AB$ indicates that the first qubit of each EPR pair is held by $A$ and the second qubit is held by $B$. In Fig.~\ref{fig:model}, the entanglement generator is located at $B$. Hence, the first qubit of the EPR pairs $\ket{\Phi^+}_{A}$ has to be sent by $B$ through the quantum channel, while the second qubit of the EPR pairs $\ket{\Phi^+}_{B}$ is available immediately at $B$. After $A$ obtains the first qubit of the EPR pairs $\ket{\Phi^+}_{A}$, it can be exploited for transmitting the quantum information embedded within the logical quantum qubit $\ket{\psi}$. In addition to the pre-shared EPR pairs, $A$ and $B$ are also connected via a classical communication channel, which is considered to be noise-free\footnote{This assumption is not restrictive since we focus our attention on the quantum noise only. In case of a realistic noisy classical channel, the well-known classical error-mitigation techniques can be implemented.}.

The main goal of the quantum communication model of Fig.~\ref{fig:model} is to faithfully transfer the quantum state $\ket{\psi}$ from $A$ to $B$ assisted by the pre-shared EPR pairs and also by classical communications. To achieve this goal, $A$ may exploit the noisy pre-shared EPR pairs $\ket{\Phi^+}_{A}$ for appropriately encoding the logical qubits $\ket{\psi}$ into $\ket{\overline{\psi}}$, which is sent to $B$. In addition to the received encoded quantum state $\ket{\tilde{\psi}}$, $B$ also obtains the classical bits gleaned from the measurement of the EPR-pair members $\ket{\Phi^+}_{A}$ at $A$. Finally, $B$ performs a decoding procedure to reconstruct the original quantum state $\ket{\psi}$ of the logical qubits by utilizing the qubits of the EPR-pair members $\ket{\Phi^+}_{B}$ at $B$.

In this treatise, we consider one of the most general quantum channel models, namely the quantum depolarizing channel $\mathcal{N}(\cdot)$, a type of quantum Pauli channel. For a single-qubit system, the quantum depolarizing channel is described by~\cite{nielsen2000quantum}
\begin{equation}
	\mathcal{N}(\rho) = (1 - p)\rho + \frac{p}{3} \left( {X}\rho{X} + {Y}\rho{Y} + {Z}\rho{Z} \right),
\label{eq:channel_single}
\end{equation} 
where $\lbrace I, X, Y, Z \rbrace$ are the Pauli matrices, $\rho$ denotes the density matrix of the input quantum state, and $p$ denotes the depolarizing probability of the quantum channel $\mathcal{N}(\cdot)$. The Kraus operators of $\mathcal{N}(\cdot)$ are given by $N_1 = \sqrt{1-p}I$, $N_2 = \sqrt{\frac{p}{3}} X$, $N_3 = \sqrt{\frac{p}{3}}Y$, $N_4=\sqrt{\frac{p}{3}}Z$~\cite{nielsen2000quantum}.


\section{Quantum Communication with Noisy Pre-Shared Entanglement}
\label{Quantum Communication with Noisy Pre-Shared Entanglement}

\begin{figure}[t]
    \center
    \includegraphics[width=\columnwidth]{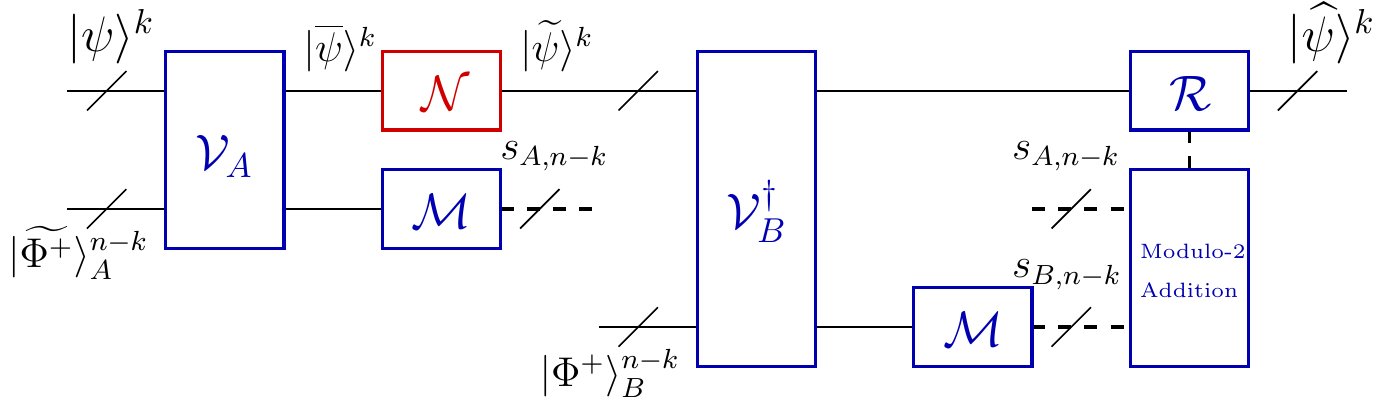}
	\caption{The scheme proposed for performing noiseless quantum communication using noisy pre-shared EPR pairs.}
	\label{fig:general}
\end{figure}

In this section, we present the general concept of our proposed scheme for performing both error-detection and error-correction. The schematic of the proposed scheme is depicted in Fig~\ref{fig:general}. Its operation commences by preparing the initialized quantum state as follows:
\begin{equation}
	\ket{\psi_p} = \ket{\psi}^k \otimes \ket{\Phi^+}_{AB}^{n-k},
	\label{eq:free-global-state}
\end{equation}
where $\ket{\psi}^k$ represents the quantum state of $k$ logical qubits, while $\ket{\Phi^+}_{AB}^{n-k}$ represents $(n-k)$ pairs of pre-shared EPR pairs $\ket{\Phi^+}$ between $A$ and $B$. The subscripts $A$ and $B$ indicate that half of the EPR pairs are held by $A$ and the other half by $B$.

As we elucidated in Section~\ref{System Model}, the generation and the distribution of the EPR pairs to $A$ are contaminated by the quantum noise imposed by the quantum channels. Let us denote the $(n-k)$-tuple Pauli operator inflicted by the quantum channel as $P_{n-k}$. Then, we have
\begin{equation}
	\ket{\widetilde{\Phi^+}}_A^{n-k} = P_{n-k}\ket{\Phi^+}_A^{n-k}.
\end{equation}
The quantum state $\ket{\psi}^k$ of the logical qubits is encoded by a quantum encoder $\mathcal{V}_A$, where we exploit the noisy EPR-pair members at $A$ $\ket{\widetilde{\Phi^+}}_A^{n-k}$. The encoded state $\ket{\overline{\psi}}^k$ of the logical qubits is then sent through the quantum channel $\mathcal{N}(\cdot)$. Let us denote the $k$-tuple Pauli operator inflicted by the quantum channel as $P_k$. Then, we have
\begin{equation}
\ket{\widetilde{\psi}}^k = P_k\ket{\overline{\psi}}^k.
\end{equation}

At the receiver side, the quantum decoder $\mathcal{V}_B^{\dagger}$ of Fig.~\ref{fig:general} decodes the corrupted quantum state $\ket{\widetilde{\psi}}^k$ with the aid of the $(n-k)$ EPR-pair members $\ket{\Phi^+}_B^{n-k}$ at $B$. To design the quantum encoder $\mathcal{V}_A$ and the quantum decoder $\mathcal{V}_B^{\dagger}$, we impose the reversible property\footnote{We note that in Fig.~\ref{fig:general} there is a little notation-abuse, since we use the symbols $\mathcal{V}_A$ and $\mathcal{V}_B^{\dagger}$ to denote the encoding and decoding performed on the qubits available at $A$ and $B$, respectively. Instead, in~\eqref{eq:reversible}, $\mathcal{V}_A$ and $\mathcal{V}_B^{\dagger}$ denote the encoder and decoder acting on the global quantum state $\ket{\psi}^k \otimes \ket{\Phi^+}_{AB}^{n-k}$. However, this notation abuse can be tolerated since $\mathcal{V}_A$ and $\mathcal{V}_B^{\dagger}$ in~\eqref{eq:reversible} leave the qubits unavailable at $A$ and $B$, respectively, unchanged.} on the initialized quantum state in~\eqref{eq:free-global-state}, which is formulated as
\begin{equation}
	\mathcal{V}_B^{\dagger}\mathcal{V}_A (\ket{\psi}^k \otimes \ket{\Phi^+}_{AB}^{n-k})  = \ket{\psi}^k \otimes \ket{\Phi^+}_{AB}^{n-k}.
\label{eq:reversible}
\end{equation}

\begin{remark}
	We note that in conventional QECCs, the reversible property of a noise-free scenario can always be guaranteed, since the quantum encoder $\mathcal{V}$ and decoder $\mathcal{V}^{\dagger}$ act on the same physical qubits. By contrast, in our scheme, the quantum encoder $\mathcal{V}_A$ only processes the logical qubits $\ket{\psi}^k$ and the EPR-pair members at $A$, whilst the quantum decoder $\mathcal{V}_B^{\dagger}$ only processes the logical qubits $\ket{\widetilde{\psi}}^k$ received via the noisy quantum channel $\mathcal{N}(\cdot)$ and the EPR-pair members at $B$.
\end{remark}

By denoting the density matrix of $\ket{\psi_p} = \ket{\psi} \otimes^k \ket{\Phi^+}_{AB}^{n-k}$ as $\overline{\rho}$, it is possible to reformulate the proposed general scheme of Fig.~\ref{fig:general} as the following supermap $\mathcal{S}$:
\begin{equation}
	\mathcal{S}(\mathcal{V}_A, \mathcal{V}_B^{\dagger},\mathcal{N}, \overline{\rho}) = \sum_{i,j} (V_B N_j V_A N_i)\overline{\rho} (V_B N_j V_A N_i)^{\dagger}.
\label{eq:supermap1}
\end{equation}
In~\eqref{eq:supermap1}, we take into account the effects of the quantum noise inflicted by the quantum channels utilized for both the distribution of the EPR-pair members at $A$ and for the transmission of the encoded state of the logical qubits. Furthermore, in~\eqref{eq:supermap1}, $N_i$, $N_j$ represent the Kraus operators of the quantum channels\footnote{To be more precise and with a little notation-abuse, $N_i$, $N_j$ denote the extended Kraus operators of the quantum channels, which account for the specific qubits affected by the quantum channels and for the increased dimension induced by the supermap of~\eqref{eq:supermap1}, acting on the global state $\ket{\psi} \otimes^k \ket{\Phi^+}_{AB}^{n-k}$.}, while $V_A$ and $V_B$ are the matrix representations of the quantum encoder and decoder, respectively.

The scheme proposed in Fig.~\ref{fig:general} is completed by local measurements $\mathcal{M}$ on the EPR pairs whose outcomes control the operator $\mathcal{R}$ depending the particular error-control strategy implemented. Specifically, to perform the associated error-control procedure, local measurements of the EPR pairs are performed for obtaining the classical bits\footnote{When $n-k$ EPR pairs are considered, the local measurements of the EPR pairs produce $2(n-k)$ outcomes. To denote the associated vectors, we utilize the notation $\underline{s}$.} $\underline{s}_{A,n-k}$ and $\underline{s}_{B,n-k}$. Since no joint measurements are applied to the EPR pairs for the sake of reducing the number of quantum channels utilization, a syndrome-like quantity may be constructed from the modulo-2 addition of the classical measurement results as follows:
\begin{equation}
	\underline{s}_{n-k} = \underline{s}_{A,n-k} \oplus \underline{s}_{B,n-k}.
\label{eq:sindrome}
\end{equation}
It is important to note that both $A$ and $B$ have chosen the appropriate pre-determined measurement basis $\mathcal{M}$ for each of the EPR pairs.

In the case of the proposed error-detection schemes, the operator $\mathcal{R}$ of Fig.~\ref{fig:general} acts as a discard-and-retain unit based on the syndrome of~\eqref{eq:sindrome}. More specifically, if the syndrome values of~\eqref{eq:sindrome} indicate the presence of errors, i.e. the syndrome values are not zeros $(\underline{s}_{n-k} \neq \underline{0})$, the operator $\mathcal{R}$ will decide to discard the logical qubits $\ket{\psi}^k$, otherwise it will retain the logical qubits. By contrast, in the case of the proposed error-correction schemes, the operator $\mathcal{R}$ represents an error-recovery procedure based on maximum-likelihood decoding relying on the syndrome values of~\eqref{eq:sindrome}. Specifically, the error-recovery procedure can be formally expressed as
\begin{equation}
	\widehat{L}_k(\underline{s}_{n-k}) = \underset{L_k}{\mathrm{arg \, max \,}}  P(L_k|\underline{s}_{n-k}), 
\end{equation}
where $ P(L_k|\underline{s}_{n-k})$ denotes the probability of experiencing the logical error $L_k$ imposed on the logical qubits $\ket{\psi}^k$, given that we obtain the syndrome values $\underline{s}_{n-k}$.


\section{Error-Detection Scheme}
\label{Error-Detection Scheme}

In this section, we consider the error-detection of either a single logical qubit or of two logical qubits and carry out its performance analysis. We rely on Definition~\ref{def:1} and~\ref{def:2} for characterizing the performance of the proposed error-detection schemes.

\begin{definition}
	The success probability $p_s$ of the proposed error-detection schemes is defined as the conditional probability of obtaining the legitimate quantum state $\rho$ of the logical qubits, given that we obtain the all-zero syndrome values $\underline{s}_{n-k} = \underline{0}$:
	\begin{equation}
		p_s = p(\rho | \underline{s}_{n-k} = \underline{0}) = \frac{p(\rho \cap \underline{s}_{n-k} = \underline{0})}{p(\underline{s}_{n-k} = \underline{0})}.
		\label{eq:success}
	\end{equation}
	The relationship between the qubit error ratio (QBER) and the success probability $p_s$ can simply be defined as $\text{QBER} = 1 - p_s$.
\label{def:1}
\end{definition}

\begin{definition}
	The yield $Y$ of the proposed error-detection schemes is defined as the ratio of $k$ logical qubits retained after the detection to the $n$ uses of the quantum channel $\mathcal{N}(\cdot)$:
	\begin{equation}
    	Y = \left( \frac{k}{n} \right) p(\underline{s}_{n} = \underline{0}).
    	\label{eq:yield}
	\end{equation}
	\label{def:2}
\end{definition}

Readers from the classical communication field may notice the relationship between the \textit{yield} and \textit{goodput} metrics. While yield has been widely used in the QED literature, goodput is a common metric utilized for normalizing the performance of classical coded communication systems with respect to the associated coding rate. The notion of goodput in the quantum domain is clarified in~\cite{chandra2019near}, where it is used for comparing the performance of various QECCs exhibiting different quantum coding rates and for determining their performance discrepancies with respect to the quantum capacity also known as the quantum hashing bound. We underline that yield and goodput are not the same metric, although they are intimately linked. More specifically, the goodput $G$ is defined as the product of the success probability of a given QECC by its quantum coding rate~\cite{chandra2019near}. Therefore, the goodput of our proposed scheme may be reformulated as in Definition~\ref{def:3}.

\begin{definition}
    The goodput $G$ of the proposed error-detection schemes is defined as the product of the success probability $p_s$ by the ratio of $k$ logical qubits to the $n$ uses of the quantum channel $\mathcal{N}(\cdot)$:
    \begin{equation}
        G = \left( \frac{k}{n} \right) p_s = \left( \frac{k}{n}\right)(1 - \text{QBER}).
        \label{eq:goodput}
    \end{equation}
	\label{def:3}
\end{definition}

By comparing Definition~\ref{def:2} and Definition~\ref{def:3} we can observe the intrinsic relationship between the yield and the goodput.

\subsection{Error-Detection for A Single Logical Qubit}
\label{Error-Detection for A Single Logical Qubit}

\begin{figure}[t]
    \center
    \includegraphics[width=\columnwidth]{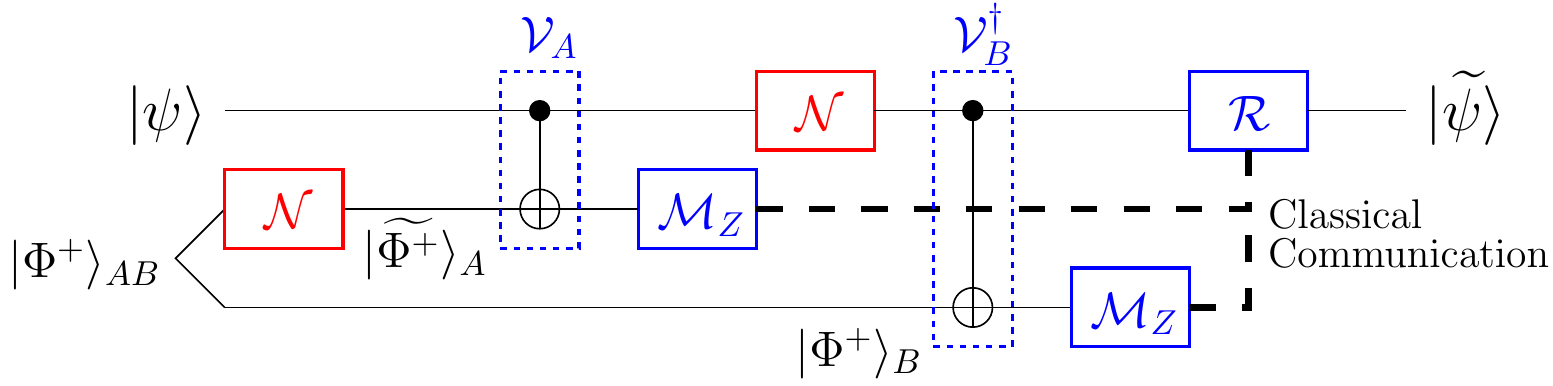}
    \caption{The quantum circuit conceived for performing a single-qubit error-detection using a single noisy EPR pair.}
    \label{fig:scheme15}
\end{figure}

Let us consider the proposed single qubit error-detection scheme depicted in Fig.~\ref{fig:scheme15}, which utilizes only a single noisy EPR pair. More specifically, the encoding and decoding circuit of Fig.~\ref{fig:general} is detailed in Fig.~\ref{fig:scheme15}. We design the quantum encoder and decoder for ensuring that the reversible condition of~\eqref{eq:reversible} is satisfied. The quantum encoder $\mathcal{V}_A$ and quantum decoder $\mathcal{V}_B^{\dagger}$ of Fig.~\ref{fig:scheme15} can be represented using unitary matrices as follows:
\begin{align}
    \label{eq:va}
    V_A = \ket{0} \bra{0} \otimes I \otimes I + \ket{1} \bra{1} \otimes X \otimes I, \nonumber \\
    V_B = \ket{0} \bra{0} \otimes I \otimes I + \ket{1} \bra{1} \otimes I \otimes X.
\end{align}
By scrutinizing~\eqref{eq:va}, it is readily seen that the reversible property is indeed satisfied, i.e. $\mathcal{V}_B^{\dagger} \mathcal{V}_A \left( \ket{\psi} \otimes \ket{\Phi^+}_{AB} \right) = \ket{\psi} \otimes \ket{\Phi^+}_{AB}$. Finally, the EPR pair is measured in the $Z$ basis $(\mathcal{M}_Z = \lbrace \ket{0} \bra{0}, \ket{1} \bra{1} \rbrace)$. The performance of the scheme proposed in Fig.~\ref{fig:scheme15} is characterized by Proposition~\ref{prop_1}.

\begin{proposition}
\label{prop_1}
	The success probability of the error-detection scheme depicted in Fig.~\ref{fig:scheme15} over quantum depolarizing channels relying on a single noisy EPR pair is given by:
\begin{align}
    p_s = 1 - \frac{2p}{3} - \frac{2p^2}{3} - \frac{8p^3}{27} + \frac{16p^4}{81} + \mathcal{O}(p^5),
\label{eq:qber1}
\end{align}
while the yield is given by:
\begin{equation}
    Y = \frac{1}{2}\left( 1-\frac{4p}{3}+\frac{8p^2}{9} \right).
\label{eq:yield1}
\end{equation}
\begin{IEEEproof}
Please refer to Appendix~A.
\end{IEEEproof}
\end{proposition}

First, we compare our proposed scheme to the state-of-the-art QED+QT schemes. Specifically, we compare the scheme proposed in Fig.~\ref{fig:scheme15} to the single-round recurrence QED of~\cite{bennett1996purification} and to the quantum stabilizer code (QSC)-based QED of~\cite{bennett1996mixed, matsumoto2003conversion, lidar2013quantum} having the stabilizer operator of $S = ZZ$. We assume that the quantum teleportation step is noise-free and therefore the QBER of the benchmark schemes is directly determined by the QBER of the associated QED scheme. Note that both the benchmark schemes require two noisy pre-shared EPR pairs, while our proposed scheme only needs a single noisy pre-shared EPR pair.

\begin{figure}[t]
\center
\includegraphics[width=\columnwidth]{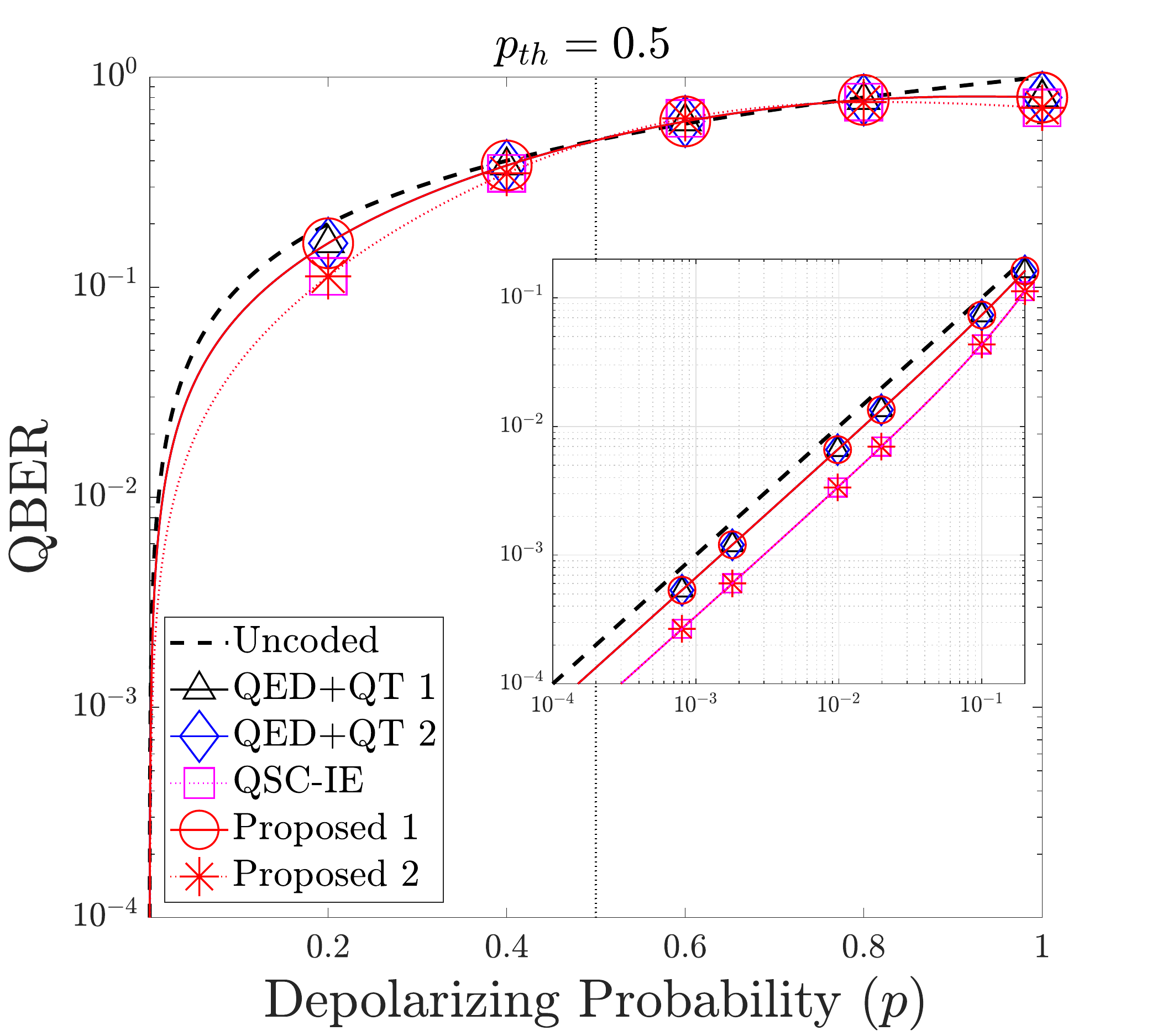}
\caption{The QBER of our proposed error-detection schemes compared to the existing schemes for mitigating the effect of quantum depolarizing channels. The uncoded QBER curve is given by $\text{QBER} = p$. The inset is the QBER relying on log-log scale.}
\label{fig:result11}
\end{figure}

\begin{figure*}[t]
    \centering
    \begin{subfigure}{0.495\linewidth}
        \includegraphics[width=\linewidth]{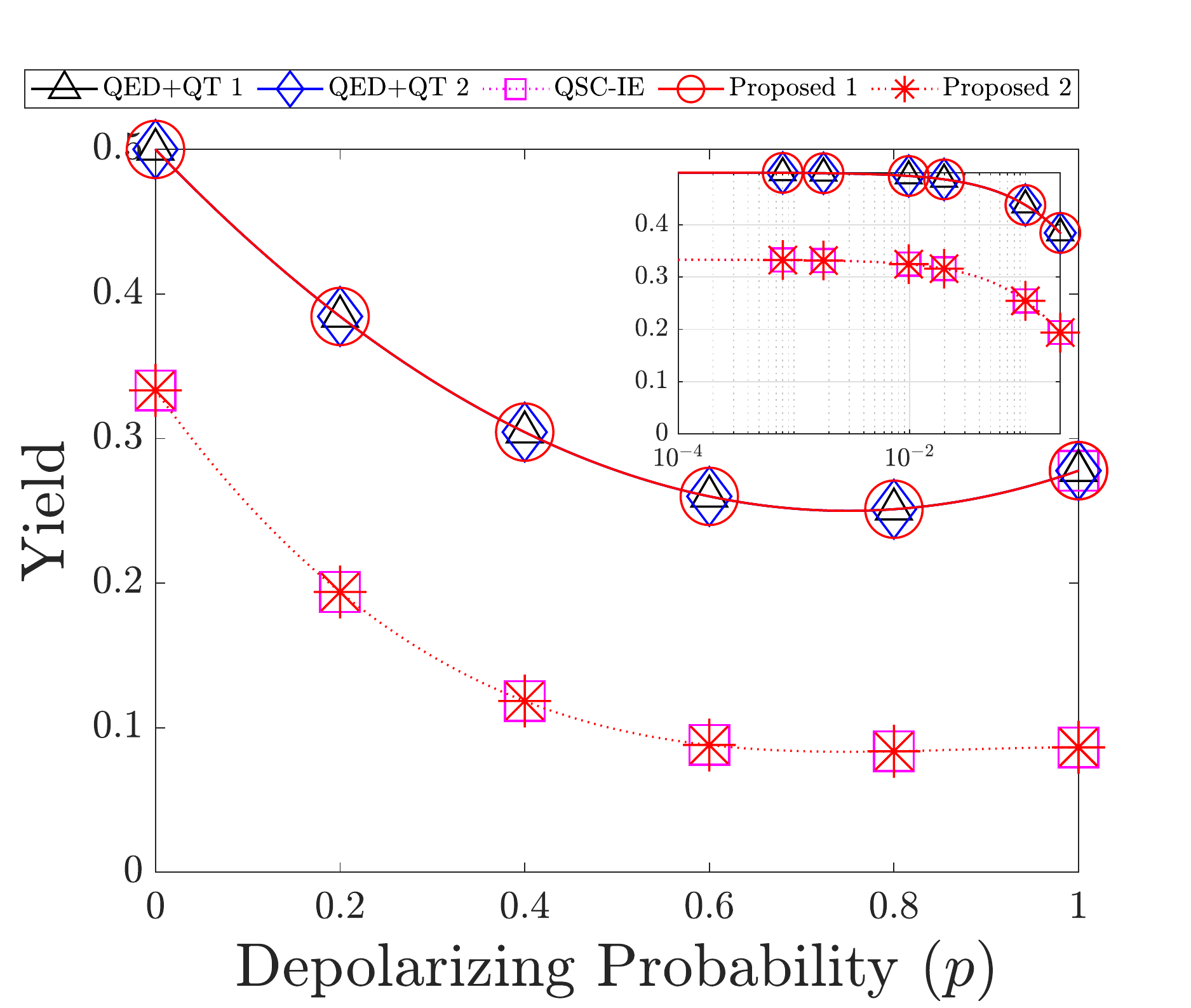}
		\caption{Yield}
		\label{fig:result07}
    \end{subfigure}
    \begin{subfigure}{0.495\linewidth}
        \includegraphics[width=\linewidth]{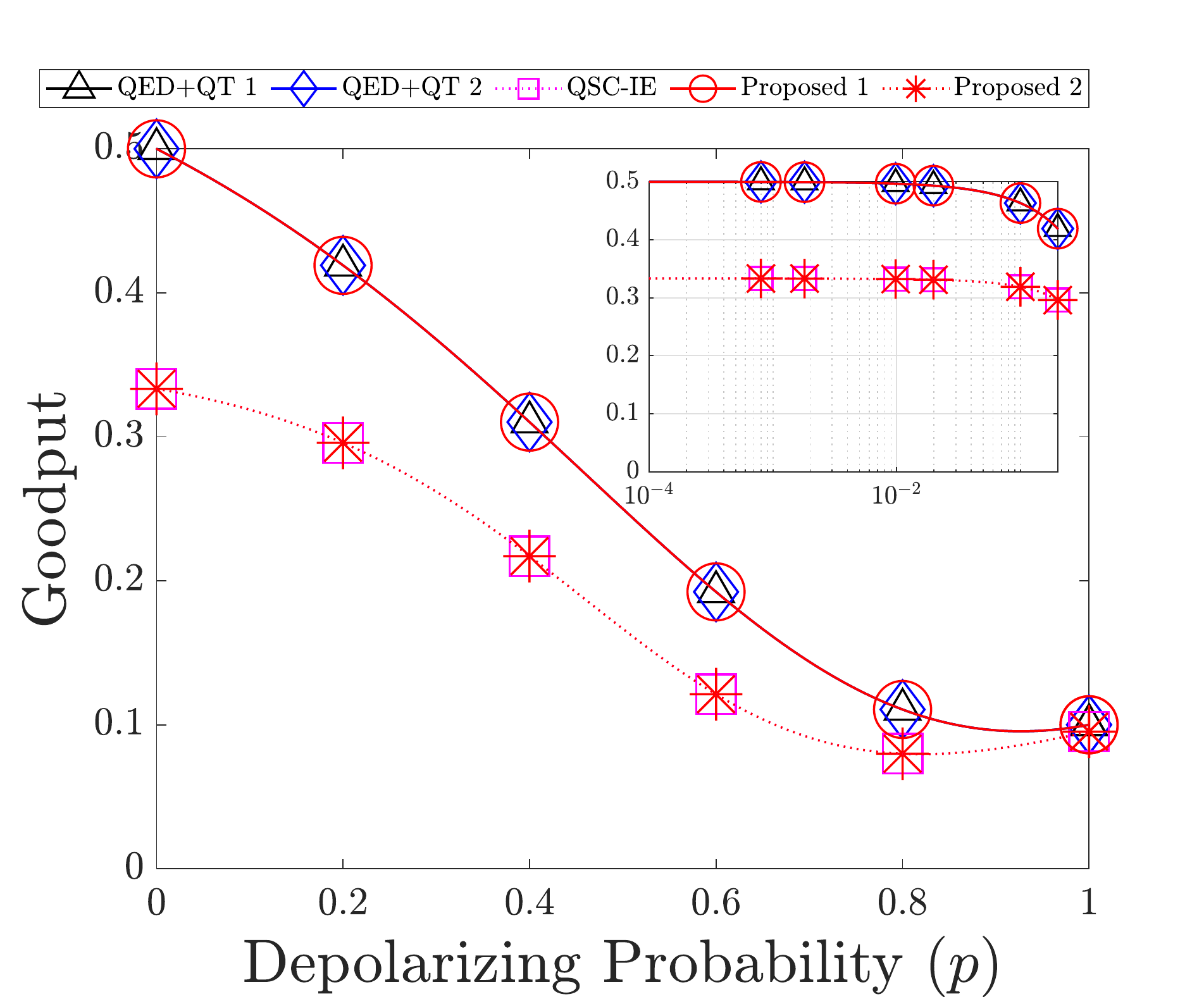}
		\caption{Goodput}
		\label{fig:goodput1}
    \end{subfigure}
\caption{The (a) yield and the (b) goodput of the proposed error-detection schemes compared to the existing schemes for mitigating the effect of quantum depolarizing channels. The insets are the yield and the goodput in the logarithmic $x$ axis.}
\label{fig:result07goodput1}
\hrulefill
\end{figure*}

The QBER is portrayed in Fig.~\ref{fig:result11}, where we label the performance of the scheme presented in Fig.~\ref{fig:scheme15} as `Proposed 1', the recurrence-based scheme as `QED+QT 1', and the QSC-based scheme as `QED+QT 2'. We observe that the QBER of the scheme presented in Fig.~\ref{fig:scheme15} matches that of QED+QT schemes, without requiring the additional quantum teleportation step, which also relies on the idealized assumption of being noise-free for both benchmarks. Furthermore, we observe that all the schemes considered are only capable of detecting a single $X$ error. Additionally, we mark the probability threshold $p_{th}$ using the vertical black dotted line in Fig.~\ref{fig:result11}, highlighting the particular depolarizing probability value, below which the proposed error-detection scheme improves the QBER of the logical qubit. Specifically, in Fig.~\ref{fig:result11}, we obtain the probability threshold of $p_{th} = 0.5$.

In Fig.~\ref{fig:result07goodput1}(a), we report the performance of our proposed scheme in terms of its yield. We observe that our proposed scheme provides an identical yield to the benchmark schemes. However, two noisy pre-shared EPR pairs are used for obtaining a single less noisy pre-shared EPR pair for both the recurrence-based and the QSC-based QED+QT schemes. This means that during the process one of the noisy pre-shared EPR pairs is discarded. By contrast, our protocol only needs a single noisy pre-shared EPR pair for achieving the same QBER performance. Finally, the goodput of our proposed error-detection scheme is presented in Fig.~\ref{fig:result07goodput1}(b), which confirms again the intrinsic relationship between the yield and the goodput. Specifically, our proposal that provides an identical yield, gives us also an identical goodput.

Apart from its benefit of utilizing fewer pre-shared EPR pairs, our proposed scheme also offers a pair of additional advantages:
\begin{itemize}
    \item It does not suffer from long communication delay, since it does not require the consecutive steps of performing QED followed by quantum teleportation.
    \item It requires fewer controlled-NOT (CNOT) quantum gates. Quantitatively, the proposed scheme of Fig.~\ref{fig:scheme15} requires a total of only two CNOT gates. By contrast, the recurrence-based QED+QT scheme requires a total of three CNOT gates: two for a single-round recurrence QED and one for quantum teleportation. As for the QSC-based QED+QT scheme, we need a total of seven CNOT gates: four for the measurement of stabilizer operators, two for the quantum inverse encoder, and one for quantum teleportation. 
\end{itemize}

Let us elaborate a little further concerning the delay imposed by each quantum communication scheme specified in the first bullet point. The quantum entanglement distillation has to be completed before the quantum teleportation can be conducted within the QED+QT scheme. Specifically, within the quantum entanglement distillation step itself, the transmission delay is imposed by the associated classical communications. Let us assume that each classical communication takes a duration of $t_c$. Therefore, for a recurrence QED scheme having $m$ rounds of distillation, the total transmission delay is equal to $mt_c$, since for each round of distillation requires a backward- and a forward-oriented classical communication phase -- both of which can be carried out simultaneously for example using wavelength division multiplexing. By contrast, the total transmission delay imposed by a QSC-based QED is simply equal to $t_c$, since it only needs a forward-oriented classical communication. Once the QED step has been completed, quantum teleportation has to be performed for transferring the quantum information from the transmitter to the receiver. Since quantum teleportation also requires another forward-oriented classical information phase, an additional delay of $t_c$ is introduced by the QED+QT scheme. Therefore, we have a total of $2t_c$ transmission delay for the QSC-based QED+QT scheme. In case of recurrence-based QED+QT scheme, we have a transmission delay of $(m+1)t_c$. By contrast, the total transmission delay is only equal to $t_c$ for both the QSC-IE and the proposed schemes. However, it is important to note that we underestimate the QSC-IE quantum information processing delay, since QSC-IE scheme utilizes the stabilizer measurements differently from the proposed scheme.

Arguably, the delay imposed by the transmission of classical information required by the QED+QT scheme can be avoided by performing QED in an \textit{asynchronous} way, implying that the QED is activated before the transmitter and the receiver have agreed to initiate their quantum communication. Consequently, an asynchronous QED+QT scheme requires a long-expiry quantum memory to store the distilled EPR pairs. However, the assumption of having a long-expiry quantum memory at both the transmitter and the receiver is indeed a strong one at the current state-of-the-art~\cite{nguyen2016exit}. In the absence of long-expiry quantum memory, an asynchronous QED+QT scheme can be employed by performing continuous QED until both the transmitter and the receiver finally decide to initiate their quantum communication. However, it is clear that continuously performing QED consumes a high number of noisy pre-shared EPR pairs during the waiting period. Therefore, in this treatise, we consider an \textit{on-demand} quantum communication model, as we have described in Section~\ref{System Model}. This model eliminates the stringent requirement of long-expiry quantum memory as well as the continuous operation of QED, which is achieved by only initializing quantum communication once both the transmitter and the receiver are ready to engage.

Regarding the number of CNOT gates mentioned in the second bullet point, it has been shown in~\cite{chandra2019quantum,cane2020mitigation}, that the number of CNOT gates provides a reasonable estimate of the severity of quantum error proliferation effects, when the realistic quantum encoder $\mathcal{V}_A$ and decoder $\mathcal{V}_B^{\dagger}$ are potentially error-infested. Specifically, in this case, the overall proliferation of quantum errors is heavily dependent on the number of two-qubit quantum gates -- exemplified by the CNOT gates. However, to fully characterize the performance of quantum communication schemes under the realistic scenario of having both noisy pre-shared entanglement and imperfect quantum gates, computer simulations are required. Thus, we will carry out this full-scale analysis in our future work.

\begin{remark}
	By invoking the simple scheme presented in Fig~\ref{fig:scheme15}, we can attain both an identical yield and a reduced delay, despite relying on a reduced number of CNOT gates compared to the benchmarks, which is achieved without degrading the QBER.
\end{remark}

\begin{figure}[t]
	\center
	\includegraphics[width=\columnwidth]{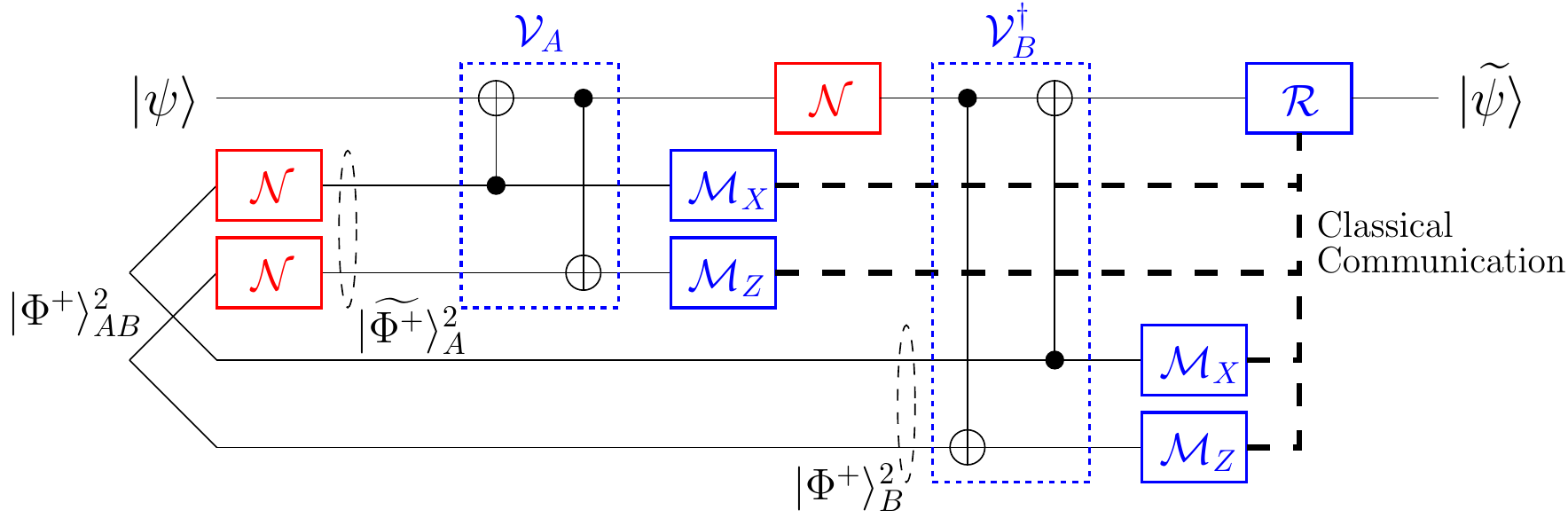}
	\caption{The quantum circuit for performing a single-qubit error-detection using two noisy pre-shared EPR pairs.}
	\label{fig:scheme16}
\end{figure}

In order to further generalize our analysis, let us compare the aforementioned schemes by using the same number of noisy pre-shared EPR-pairs. More specifically, we assume having two noisy pre-shared EPR-pairs for all the QED+QT schemes considered. Specifically, we modify the scheme proposed in Fig.~\ref{fig:scheme15} as seen in Fig.~\ref{fig:scheme16}, where the first EPR pair is measured in the $Z$ basis $(\mathcal{M}_Z = \lbrace \ket{0} \bra{0}, \ket{1} \bra{1} \rbrace)$, while the second pair in the $X$ basis $(\mathcal{M}_X = \lbrace \ket{+} \bra{+}, \ket{-} \bra{-} \rbrace)$. Let us distinguish the components of the syndrome vector in~\eqref{eq:sindrome} according to the observation basis used for the measurement. Specifically, let us denote the syndrome component obtained when the first EPR pair is measured in the $Z$ basis by $s_Z = s_A \oplus s_B$ and that obtained when the second EPR pair is measured in the $X$ basis by $s_X = s_A \oplus s_B$. The operator $\mathcal{R}$ acts as follows: if $s_Z = 0$, the measurement of the second EPR pair is performed to obtain $s_X$. Otherwise, the logical qubit is discarded immediately, since there is no need to measure the syndrome value $s_X$, if the syndrome value $s_Z$ already indicates that the logical qubit is corrupted. The aforementioned decision strategy is summarized as a look-up table (LUT) in Table~\ref{table:proposed23}(a). The performance of the error-detection scheme depicted in Fig.~\ref{fig:scheme16} is quantified in terms of its QBER and yield presented in Proposition~\ref{prop_2}.

\begin{table}[b]
\caption{Syndrome values and associated decision $\mathcal{R}$ for the error-detection schemes.}
\label{table:proposed23}
    \begin{subtable}[b]{0.5\textwidth}
	\centering
    \caption{Scheme in Fig.~\ref{fig:scheme16}.}
    \label{table:proposed2}
	\begin{tabular}{|c|c|c|}
	\hline
	$s_Z$ & $s_X$ & Decision $\mathcal{R}$ \\
	\hline
	0 & 0 & Retain \\
	\hline
	0 & 1 & Discard \\
	\hline
	1 & n.a. & Discard \\
	\hline
	\end{tabular}
    \end{subtable}
    \begin{subtable}[b]{0.5\textwidth}
  	\vspace{5mm}
    \centering
    \caption{Scheme in Fig.~\ref{fig:scheme13}.}
    \label{table:proposed3}
	\begin{tabular}{|c|c|c|}
	\hline
	$s_X$ & $s_Z$ & Decision $\mathcal{R}$ \\
	\hline
	0 & 0 & Retain \\
	\hline
	0 & 1 & Discard \\
	\hline
	1 & 0 & Discard \\
	\hline
	1 & 1 & Discard \\
	\hline
	\end{tabular}
    \end{subtable}
\end{table}

\begin{proposition}
\label{prop_2}
	The success probability of the proposed error-detection scheme of Fig.~\ref{fig:scheme16} operating over quantum depolarizing channels by utilizing two noisy EPR pairs is:
\begin{equation}
	p_s = 1 - \frac{p}{3} - \frac{8p^2}{9} - \frac{32p^3}{27} - \frac{64p^4}{81} + \mathcal{O}(p^5),
\label{eq:p_s_prop_2}
\end{equation}
while the yield is expressed as
\begin{equation}
	Y = \frac{1}{3} \left( 1 - \frac{8p}{3} + \frac{28p^2}{9} - \frac{32p^3}{27} \right).
\label{eq:y_prop_2}
\end{equation}
\begin{IEEEproof}
Please refer to Appendix~B.
\end{IEEEproof}
\end{proposition}

We also compare our proposed scheme to the state-of-the-art QSC-IE scheme. Specifically, we compare the scheme proposed in Fig.~\ref{fig:scheme16} to the QSC-IE scheme of~\cite{lai2012entanglement} having the stabilizer operators of $S_1 = IXX$ and $S_2 = ZZZ$. The QBER, the yield, as well as the goodput of the proposed scheme in Fig.~\ref{fig:scheme16} are portrayed in Fig.~\ref{fig:result11},~\ref{fig:result07goodput1}(a), and~\ref{fig:result07goodput1}(b), respectively, where it is labeled as `Proposed 2', while the QSC-IE benchmark scheme is labeled as `QSC-IE'. Observe in Fig.~\ref{fig:result11} that the QBER of the error-detection scheme in Fig.~\ref{fig:scheme16} outperforms all the QED+QT benchmark schemes, while providing an identical QBER to the QSC-IE benchmark scheme. We also obtain the probability threshold of $p_{th} = 0.5$ for the proposed error-detection scheme in Fig.~\ref{fig:scheme16}. However, the QSC-IE scheme only requires one pre-shared EPR pair, while our proposed scheme requires two pre-shared EPR pairs as shown in Fig.~\ref{fig:scheme16}. Nonetheless, our proposed scheme requires fewer CNOT gates and the same number of quantum channel uses by avoiding the need of stabilizer measurements. More specifically, the scheme presented in Fig.~\ref{fig:scheme16} requires a total of only four CNOT gates, while the QSC-IE scheme requires a total of eight CNOT gates: one for quantum encoder, five for the measurement of stabilizer operators, and two for the quantum inverse encoder.

\begin{remark}
	By maintaining the same maximal yield and goodput as the state-of-the-art schemes, our proposed scheme provides identical error-detection performance despite utilizing fewer CNOT gates.
\end{remark}

\subsection{Error-Detection for Two Logical Qubits}
\label{Error-Detection for Two Logical Qubits}

\begin{figure}[t]
\center
\includegraphics[width=\columnwidth]{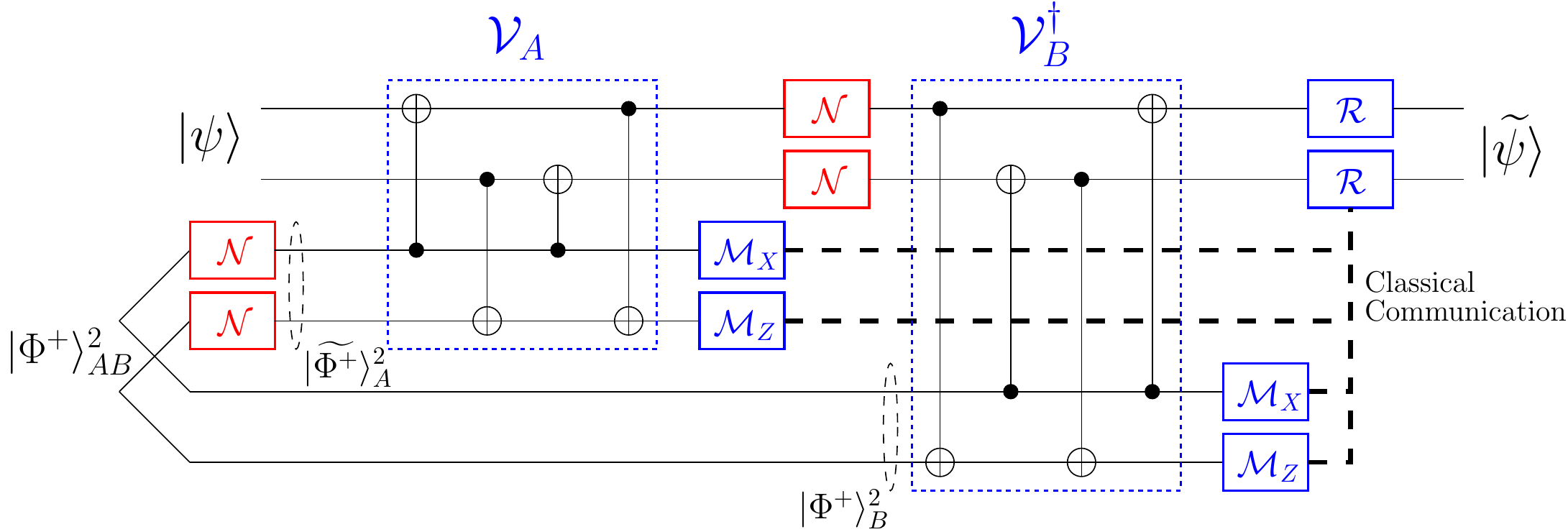}
\caption{The quantum circuit designed for the proposed error-detection for two logical qubits, which utilizes two noisy pre-shared EPR pairs.}
\label{fig:scheme13}
\end{figure}

Let us now shift our focus to the scheme presented in Fig.~\ref{fig:scheme13}, where we use two noisy EPR pairs for constructing an error-detection scheme for two logical qubits. Again, the quantum encoder $\mathcal{V}_A$ and decoder $\mathcal{V}^{\dagger}_B$ are designed for satisfying the reversible property. The resultants quantum encoder $\mathcal{V}_A$ and decoder $\mathcal{V}^{\dagger}_B$ are seen in Fig.~\ref{fig:scheme13}. The first EPR pair is measured in the $X$ basis $(\mathcal{M}_X = \lbrace \ket{+} \bra{+}, \ket{-} \bra{-} \rbrace)$, while the second pair in the $Z$ basis $(\mathcal{M}_Z = \lbrace \ket{0} \bra{0}, \ket{1} \bra{1} \rbrace)$. Additionally, the decision block $\mathcal{R}$ of Fig.~\ref{fig:scheme13} is represented by the LUT of Table~\ref{table:proposed23}(b). We summarize the performance results in Proposition~\ref{prop_2qubit_dep}.

\begin{proposition}
\label{prop_2qubit_dep}
The success probability of the proposed error-detection scheme of Fig.~\ref{fig:scheme13} operating over quantum depolarizing channels is given by:
\begin{equation}
	p_s = 1 - 2p^2 - \frac{44p^3}{9} - \frac{44p^4}{9} + \mathcal{O}(p^5),
\end{equation}
while the yield is expressed as
\begin{equation}
	Y = \frac{1}{2}\left( 1 - 4p + 8p^2 - \frac{64p^3}{9} + \frac{64p^4}{27} \right).
\end{equation}
\begin{IEEEproof}
Please refer to Appendix~C.
\end{IEEEproof}
\end{proposition}

\begin{figure*}[t]
    \centering
    \begin{subfigure}{0.495\linewidth}
        \includegraphics[width=\linewidth]{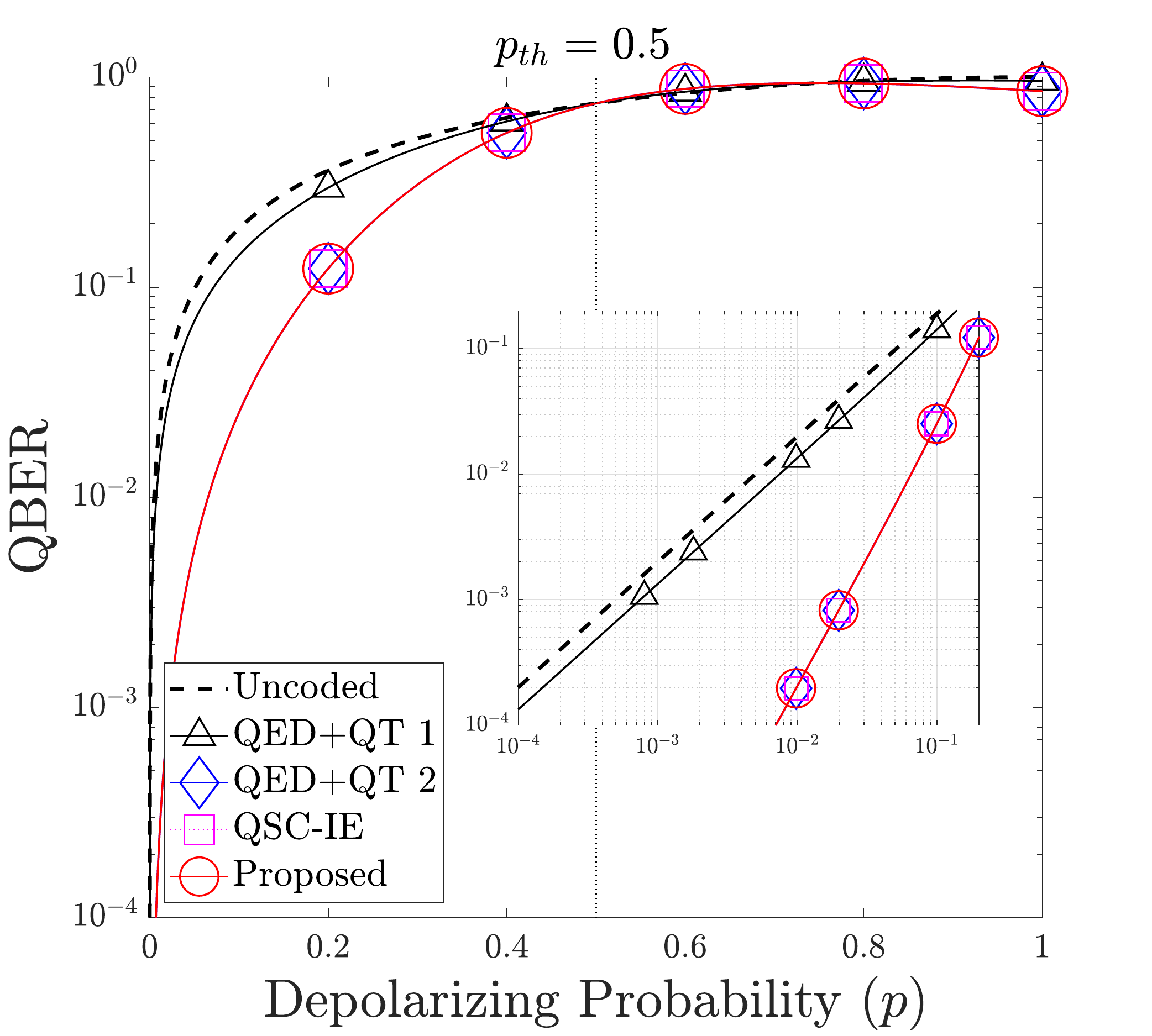}
		\caption{QBER}
		\label{fig:result12}
    \end{subfigure}
    \begin{subfigure}{0.495\linewidth}
        \includegraphics[width=\linewidth]{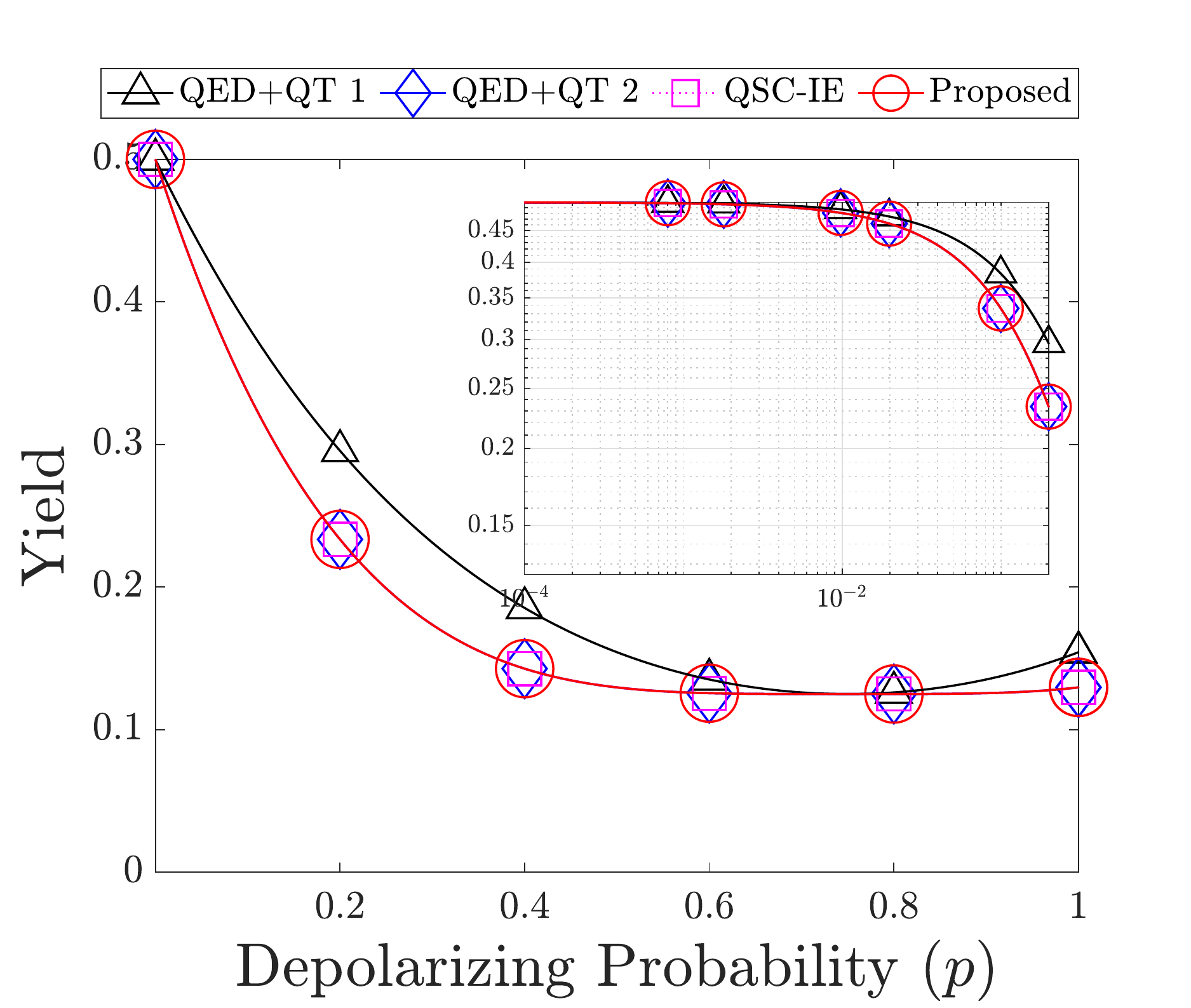}
		\caption{Yield}
		\label{fig:result08}
    \end{subfigure}
\caption{The (a) QBER and the (b) yield of the proposed error-detection scheme in Fig.~\ref{fig:scheme13} compared to the existing schemes for mitigating the effect of quantum depolarizing channels. The uncoded QBER curve is given by $\text{QBER} = 1-(1-p)^2 = 2p-p^2$. The insets are the QBER and the yield in the logarithmic $x$ axis.}
\label{fig:result1208}
\hrulefill
\end{figure*}

To benchmark the performance of the proposed scheme, we have chosen the following QED+QT schemes. Firstly, for the recurrence-based QED+QT scheme (QED+QT 1), we carry out two single-round distillations to obtain two less noisy EPR pairs from four noisy EPR pairs. Secondly, for the QSC-based QED+QT scheme (QED+QT 2), we choose the stabilizer operators of $S_1 = XXXX$ and $S_2 = ZZZZ$ to apply error-detection to a set of four noisy EPR pairs. Finally, we also include the QSC-IE scheme having the stabilizer operators of $S_1 = XXXX$ and $S_2 = ZZZZ$ as our benchmark. The uncoded QBER is given by $\text{QBER} = 1 - (1-p)^2 = 2p-p^2$, which means that any error experienced by any logical qubit within the two qubits is considered as an error. The resultant QBER is portrayed in Fig.~\ref{fig:result1208}(a), while the yield is quantified in Fig.~\ref{fig:result1208}(b).

Let us evaluate the QBER and the yield of the recurrence-based QED+QT scheme by considering a pair of identical error-detection schemes based on Fig.~\ref{fig:scheme15}. We can determine the success probability of this arrangement by taking the square of~\eqref{eq:qber1} of Proposition~\ref{prop_1}, since the legitimate quantum state of the logical qubit is retained only when both error-detection schemes make the correct decision. Consequently, the success probability $p_s$ is given by
\begin{align}
	p_s &= \left[ \frac{1-2p+\frac{10p^2}{9}}{1-\frac{4p}{3}+\frac{8p^2}{9}} \right]^2 \nonumber \\
	&= 1 - \frac{4p}{3} - \frac{8p^2}{9} + \frac{8p^3}{27} + \frac{100p^4}{81} + \mathcal{O}(p^5).
	\label{eq:qber-double}
\end{align}
Similarly, the yield $Y$ can be obtained by taking the square of $p(s_z = 0)$ of~\eqref{eq:yield1} in Proposition~\ref{prop_1} and then by normalizing it by $k/n$, where we obtain
\begin{align}
	Y &= \frac{1}{2}\left( 1-\frac{4p}{3}+\frac{8p^2}{9} \right)^2 \nonumber \\ 
	&= \frac{1}{2}\left( 1 - \frac{8p}{3} + \frac{32p^2}{9} - \frac{64p^3}{27} + \frac{64p^4}{81} \right).
	\label{eq:yield-double}
\end{align}
The QBER and yield results in~\eqref{eq:qber-double} and~\eqref{eq:yield-double} are depicted in Fig.~\ref{fig:result1208}(a) and~\ref{fig:result1208}(b), respectively, as `QED+QT 1'.

In Fig.~\ref{fig:result1208}(a), our proposed scheme outperforms the recurrence-based QED+QT scheme (QED+QT 1) for $p < 0.5$, while exhibiting an identical QBER to the QSC-based QED+QT scheme (QED+QT 2) and QSC-IE. Furthermore, we also observe the probability threshold of $p_{th} = 0.5$, which is portrayed using vertical black dotted line, for the proposed error-detection scheme in Fig.~\ref{fig:scheme13}. However, observe in Fig.~\ref{fig:result1208}(b) that the recurrence-based QED+QT scheme attains a better yield. The reason is that the recurrence-based QED+QT scheme exhibits a weaker error-detection capability than the other schemes. More specifically, each round of recurrence QED is only capable of detecting a single $X$ error. By contrast, for the QSC-based QED+QT, QSC-IE, and also our proposed schemes, they are all capable of detecting a single $X$ error and also a single $Z$ error. Consequently, the recurrence-based QED+QT scheme often makes the wrong decision of retaining the erroneous logical qubits, instead of discarding them, which is reflected in higher QBER result. However, these QBER and yield results can be achieved by utilizing fewer CNOT gates.

\begin{figure}[t]
	\center
	\includegraphics[width=\linewidth]{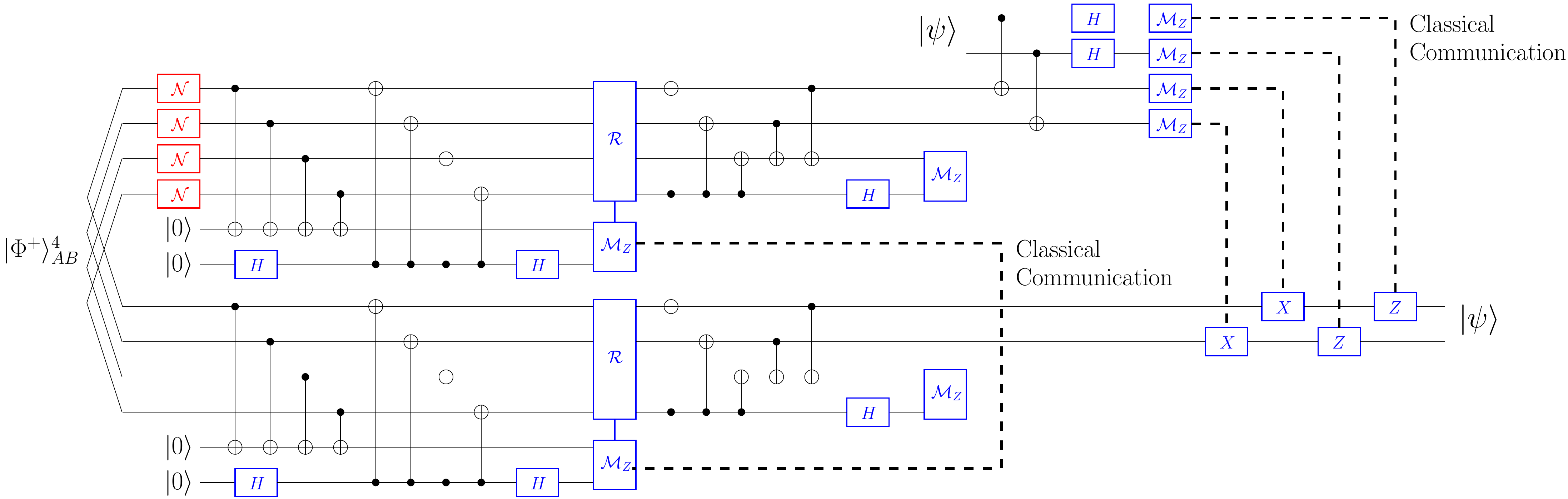}
	\caption{The quantum circuit designed for performing QSC-based QED+QT scheme using the stabilizer operators of $\mathcal{C}[4,2,2]$ code followed by a two-qubit quantum teleportation.}
	\label{fig:scheme19}
\end{figure}

\begin{figure}[t]
	\center
	\includegraphics[width=\linewidth]{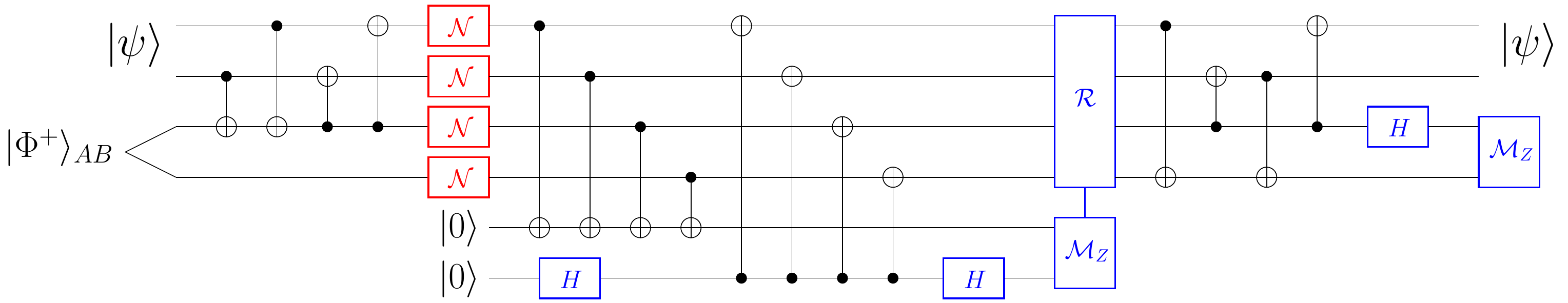}
	\caption{The quantum circuit designed for performing QSC-IE scheme using the stabilizer operators of $\mathcal{C}[4,2,2]$ code.}
	\label{fig:scheme20}
\end{figure}

Additionally, as shown in Fig.~\ref{fig:scheme13}, the total number of CNOT gates required by the entire proposed error-detection scheme is eight. As a comparison, the QSC-based QED+QT scheme requires a total of 28 CNOT gates, namely 16 for the stabilizer measurements, 10 for the quantum inverse encoder, and two for quantum teleportation. The quantum circuit of the QSC-based QED+QT scheme is portrayed in Fig.~\ref{fig:scheme19}. Meanwhile, the QSC-IE scheme, whose quantum circuit is portrayed in Fig.~\ref{fig:scheme20}, requires a total of 16 CNOT gates, namely four for the quantum encoder, eight for the stabilizer measurements, and four for the quantum inverse encoder. Therefore, our proposed scheme requires significantly fewer CNOT gates while offering an identical QBER and yield.

We summarize all the resources required by the various quantum communication schemes of Fig.~\ref{fig:result1208} in Table~\ref{table:result1}. All the schemes considered in Table~\ref{table:result1}  -- except for the QED+QT 1 scheme -- attain identical QBER and yield. To achieve this, our proposed scheme requires fewer pre-shared EPR pairs, fewer classical channel uses, and fewer CNOT gates compared to the QSC-based QED+QT scheme. By contrast, with the same number of quantum channel uses, our proposed scheme requires more pre-shared EPR pairs and more classical channel uses than the QSC-IE scheme, while utilizing fewer CNOT gates.

\begin{table*}
	\caption{Comparison of our proposed error-detection scheme with the existing approaches.}
	\small
	\centering
	\begin{tabular}{|c|r|r|r|r|r|}
		\hline
		\multirow{2}{*}{Scheme} & Quantum channel use & Logical qubits & EPR pairs & Classical channel use & \multirow{2}{*}{CNOT} \\
 		& \multicolumn{1}{c|}{$n$} & \multicolumn{1}{c|}{$k$} & \multicolumn{1}{c|}{$e$} & \multicolumn{1}{c|}{$c$} & \\
		\hline
		\hline
		QED+QT 1 & 4 & 2 & 4 & 6 & 4 \\
		\hline
		QED+QT 2 & 4 & 2 & 4 & 6 & 28 \\
		\hline
		QSC-IE & 4 & 2 & 1 & 0 & 16 \\
		\hline
		Proposed & 4 & 2 & 2 & 2 & 8 \\
		\hline
	\end{tabular}
	\label{table:result1}
\end{table*}

\begin{remark}
	While providing an identical QBER and yield to the QSC-based QED+QT and QSC-IE schemes, our error-detection scheme always requires fewer CNOT gates.
\end{remark}


\section{Error-Correction Scheme}
\label{Error-Correction Scheme}

Error-detection schemes provide dynamic yields, since they rely on a discard-and-retain action of the operator $\mathcal{R}$, while error-correction schemes provide a constant yield, since they attempt to recover the legitimate quantum state of the logical qubits from the received encoded state. Therefore, a modification of Definition~\ref{def:1} and~\ref{def:2} is required in order to accurately evaluate the performance of the proposed error-correction scheme.

\begin{definition}
	The success probability $p_s$ of the proposed error-correction scheme is defined as the sum of the conditional probabilities $p(\widehat{L}_k = L_k| \underline{s}_{n-k})$, i.e. the sum of the probabilities that the error-recovery operator $\mathcal{R}$ successfully applies $\widehat{L}_k = L_k$ based on the syndrome value $\underline{s}_{n-k}$:
	\begin{equation}
		p_s = \sum_{L_k} p(\widehat{L}_k = L_k|\underline{s}_{n-k}),
	\end{equation}
where the relationship between $p_s$ and the QBER can be expressed as $\text{QBER} = 1 - p_s$.
\label{def_1_bis}
\end{definition}

\begin{definition}
\label{def_2_bis}
	The yield $Y$ of the proposed error-correction scheme is defined as the ratio of $k$ logical qubits to the $n$ uses of the quantum channel $\mathcal{N}(\cdot)$:
	\begin{equation}
		Y = \frac{k}{n},
	\end{equation}
while its goodput is similarly defined to Definition~\ref{def:3}.
\end{definition}

\begin{figure*}[t]
\center
\includegraphics[width=\linewidth]{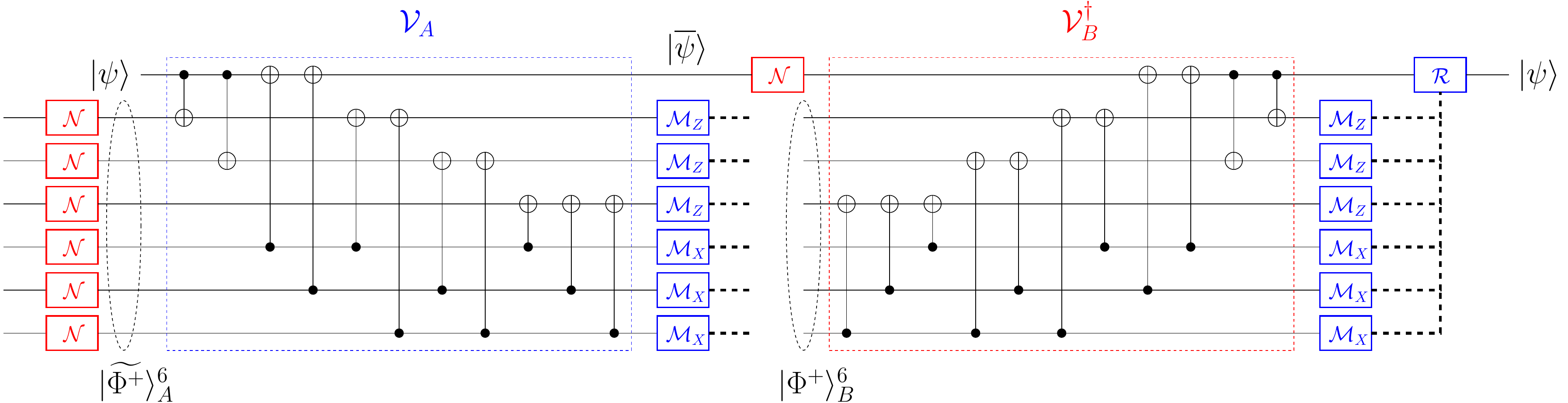}
\caption{The quantum encoder $\mathcal{V}_A$ and the quantum decoder $\mathcal{V}_B^{\dagger}$ for performing the proposed error-correction scheme.}
\label{fig:encoder}
\hrulefill
\end{figure*}

Let us now consider the quantum encoder $\mathcal{V}_A$ and decoder $\mathcal{V}_B^{\dagger}$ of Fig.~\ref{fig:encoder}. To investigate its error-correction performance, we have to check first that the scheme of Fig.~\ref{fig:encoder} is capable of discriminating all the single-qubit error patterns based on the measured syndrome values. In Fig.~\ref{fig:encoder}, we can observe that the overall scheme requires six noisy pre-shared EPR pairs, which means that we have a six-bit syndrome string denoted by $\underline{s} = s_1s_2s_3s_4s_5s_6$, where the indices $i \in \lbrace 1,2,3,4,5,6 \rbrace$ represent the EPR pair starting from the top. Therefore, for each of the single-qubit error patterns, we can evaluate the syndrome string and the associated error recovery operator, as shown in Table~\ref{table:proposed4}. Observe that the first three elements of the syndrome string $\underline{s}_Z = s_1s_2s_3$ are exclusively used for identifying $X$ errors, which are obtained from $Z$ basis measurements $(\mathcal{M}_Z = \lbrace \ket{0} \bra{0}, \ket{1} \bra{1} \rbrace)$. By contrast, the last three elements $\underline{s}_X = s_4s_5s_6$ are used for identifying $Z$ errors, which are obtained from $X$ basis measurements $(\mathcal{M}_X = \lbrace \ket{+} \bra{+}, \ket{-} \bra{-} \rbrace)$. Finally, the $Y$ errors can be identified based on the combination of $\underline{s}_Z$ and $\underline{s}_X$.

\begin{table*}
\small
\caption{Syndrome values and the associated error-recovery operator $\mathcal{R}$ of the error-correction scheme presented in Fig.~\ref{fig:encoder}.\label{table:proposed4}}
\centering
\begin{tabular}{|c|c|c||c|c|c|}
\hline
Error pattern & Syndrome & Error recovery & Error pattern & Syndrome & Error recovery \\
$P_k,P_{n-k}$ & $\underline{s}$ & $\mathcal{R}$ & $P_k,P_{n-k}$ & $\underline{s}$ & $\mathcal{R}$ \\
\hline
\hline
$XIIIIII$ & $(1 \ 1 \ 0 \ 0 \ 0 \ 0)$ & $X$ & $ZIIIIII$ & $(0 \ 0 \ 0 \ 1 \ 1 \ 0)$ & $Z$ \\
\hline
$IXIIIII$ & $(1 \ 0 \ 0 \ 0 \ 0 \ 0)$ & $I$ & $IZIIIII$ & $(0 \ 0 \ 0 \ 1 \ 0 \ 1)$ & $Z$ \\
\hline
$IIXIIII$ & $(0 \ 1 \ 0 \ 0 \ 0 \ 0)$ & $I$ & $IIZIIII$ & $(0 \ 0 \ 0 \ 0 \ 1 \ 1)$ & $Z$ \\
\hline
$IIIXIII$ & $(0 \ 0 \ 1 \ 0 \ 0 \ 0)$ & $I$ & $IIIZIII$ & $(0 \ 0 \ 0 \ 1 \ 1 \ 1)$ & $I$ \\
\hline
$IIIIXII$ & $(0 \ 1 \ 1 \ 0 \ 0 \ 0)$ & $X$ & $IIIIZII$ & $(0 \ 0 \ 0 \ 1 \ 0 \ 0)$ & $I$ \\
\hline
$IIIIIXI$ & $(1 \ 0 \ 1 \ 0 \ 0 \ 0)$ & $X$ & $IIIIIZI$ & $(0 \ 0 \ 0 \ 0 \ 1 \ 0)$ & $I$ \\
\hline
$IIIIIIX$ & $(1 \ 1 \ 1 \ 0 \ 0 \ 0)$ & $I$ & $IIIIIIZ$ & $(0 \ 0 \ 0 \ 0 \ 0 \ 1)$ & $I$ \\
\hline
\end{tabular}
\end{table*}

For the quantum depolarizing channel, we have a total of $4^7 = 16,384$ error patterns represented by the total number of combinations in terms of bit-flip $(X)$, phase-flip $(Z)$, as well as simultaneous bit-flip and phase-flip $(Y)$ errors, where we observe a total of $4,096$ correctable error patterns. After scrutinizing all $4,096$ error patterns, we obtain the Pauli weight distribution of the error patterns in quantum depolarizing channels as follows: one error pattern is the all-identity operator (weight = $0$); 21 error patterns having weight = 1; 42 error patterns having weight = 2; 252 error patterns having weight = 3; 609 error patterns having weight = 4; 1281 having weight = 5; 1428 error patterns having weight = 6; 462 error patterns having weight = 7. This distribution is identical to that of a QSC-based QED+QT scheme utilizing the stabilizer operators of the Steane code. Given that we have $p_x = p_z = p_y = \frac{p}{3}$, the success probability of the proposed error-correction scheme of Fig.~\ref{fig:encoder} in quantum depolarizing channels is given by
\begin{align}
	p_s &= \sum_{i = 0}^{n} W_i \left( \frac{p}{3} \right)^i (1-p)^{(n -i)} \nonumber \\
 &= 1 - \frac{49p^2}{3} + 56p^3 - \frac{2380p^4}{27} \nonumber \\
 &+ \frac{6160p^5}{81} - \frac{8512p^6}{243} + \frac{4824p^7}{729},
	\label{eq:correct}
\end{align}
where $W = \lbrace W_0, W_1, W_2, W_3, W_4, W_5, W_6, W_7 \rbrace = \lbrace 1, 21, 42, 252, 609, 1281, 1428, 462 \rbrace$ is the Pauli weight of the correctable error patterns. Notice that our proposed scheme is capable of correcting not only the error patterns exhibiting a Pauli weight = 1, but also several error patterns having higher Pauli weights. This is due to the degeneracy property of quantum information inherited by QECCs. Naturally, by exploiting the degeneracy property, the QBER of quantum error-control schemes, including our proposed schemes, can be improved.

\begin{figure*}[t]
    \centering
    \begin{subfigure}{0.495\linewidth}
        \includegraphics[width=\linewidth]{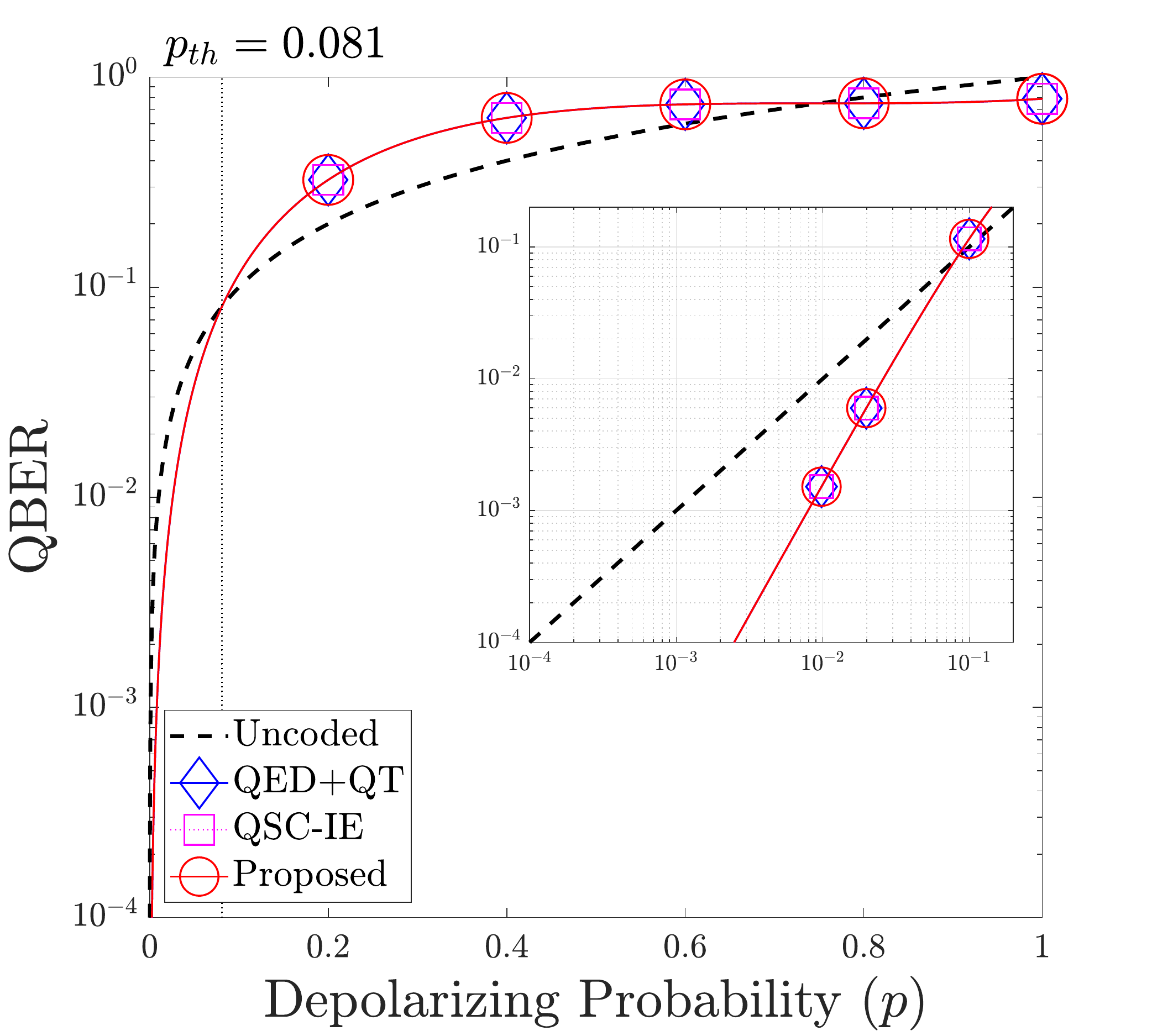}
		\caption{QBER}
		\label{fig:result23}
    \end{subfigure}
    \begin{subfigure}{0.495\linewidth}
        \includegraphics[width=\linewidth]{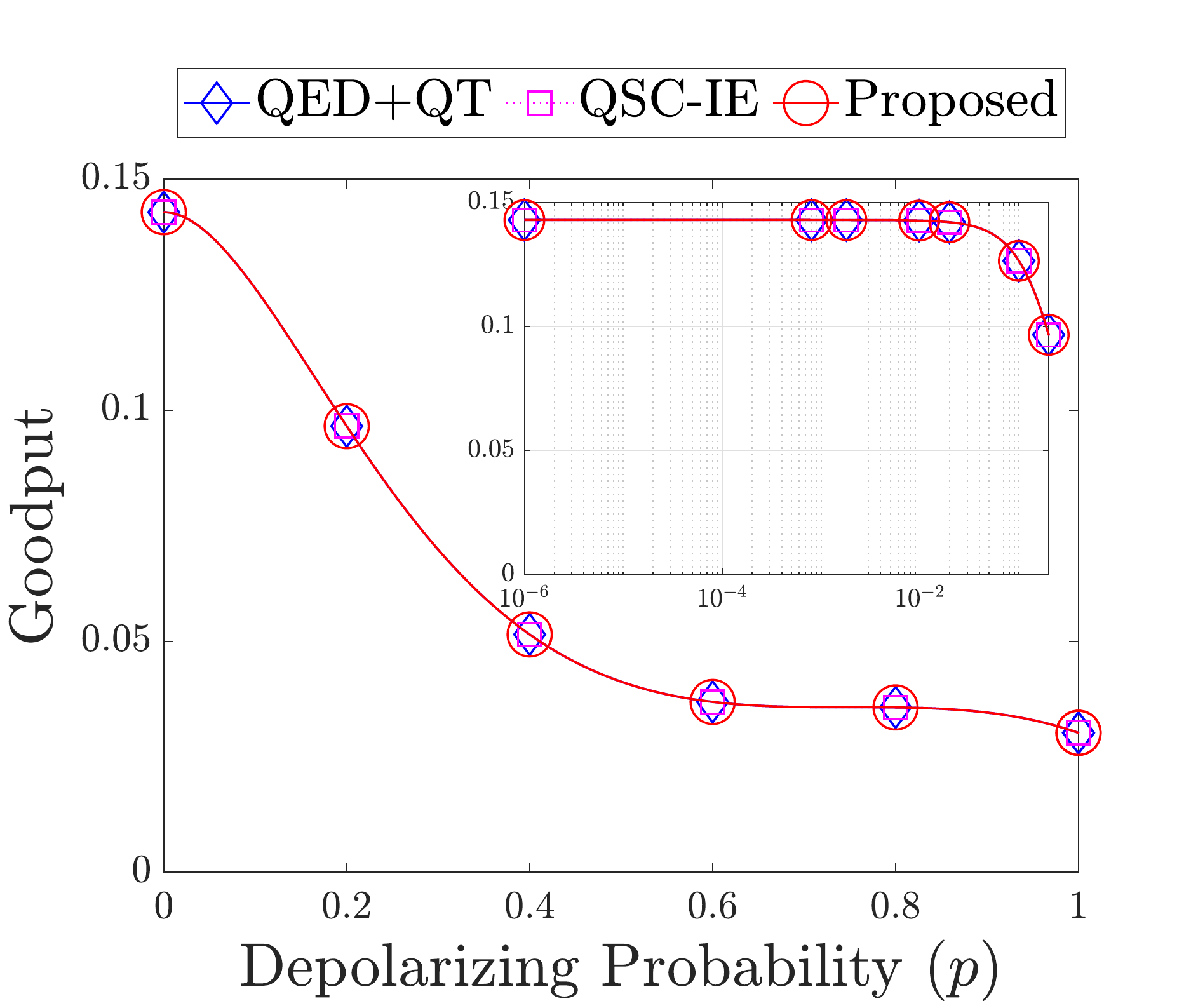}
		\caption{Goodput}
		\label{fig:goodput2}
    \end{subfigure}
	\caption{The (a) QBER and the (b) goodput of our error-correction scheme proposed in Fig.~\ref{fig:encoder} compared to the existing QSC-based scheme for mitigating the effect of quantum depolarizing channels. The insets are the QBER and the goodput in the logarithmic $x$ axis.}
	\label{fig:result23goodput2}
	\hrulefill
\end{figure*}

Let us now compare the QBER of our proposed scheme to those of the QSC-based QED+QT and QSC-IE scheme. Indeed for a fair comparison, we do not consider the recurrence-based QED+QT scheme, since it is an error-detection scheme, not an error-correction one. For the QSC-based QSC+QT scheme, we utilized the stabilizer operators of Steane code~\cite{steane1996multiple, lidar2013quantum} over seven noisy pre-shared EPR pairs. By contrast, for the QSC-IE scheme, we also utilized the stabilizer operators of Steane code, but only by using three noisy pre-shared EPR pairs~\cite{lai2012entanglement}. The resultant QBER of the proposed scheme is depicted in Fig.~\ref{fig:result23goodput2}(a), which is identical to the QBER of the QSC-based QED+QT and QSC-IE schemes employing the stabilizer operators of the Steane code. Here, we obtain the probability threshold of $p_{th} = 0.081$ for the proposed error-correction scheme in Fig.~\ref{fig:encoder}, which is indicated by the vertical black dotted line in Fig.~\ref{fig:result23goodput2}(a). The proposed, the QSC-based QED+QT, and the QSC-IE schemes all provide a yield of $Y = \frac{1}{7}$, since they perform error-correction, instead of error-detection. Consequently, as reported in Fig.~\ref{fig:result23goodput2}(b), the proposed error-correction scheme also provides an identical goodput to the QSC-based QED+QT and QSC-IE schemes.

\begin{figure*}[t]
	\center
	\includegraphics[width=\linewidth]{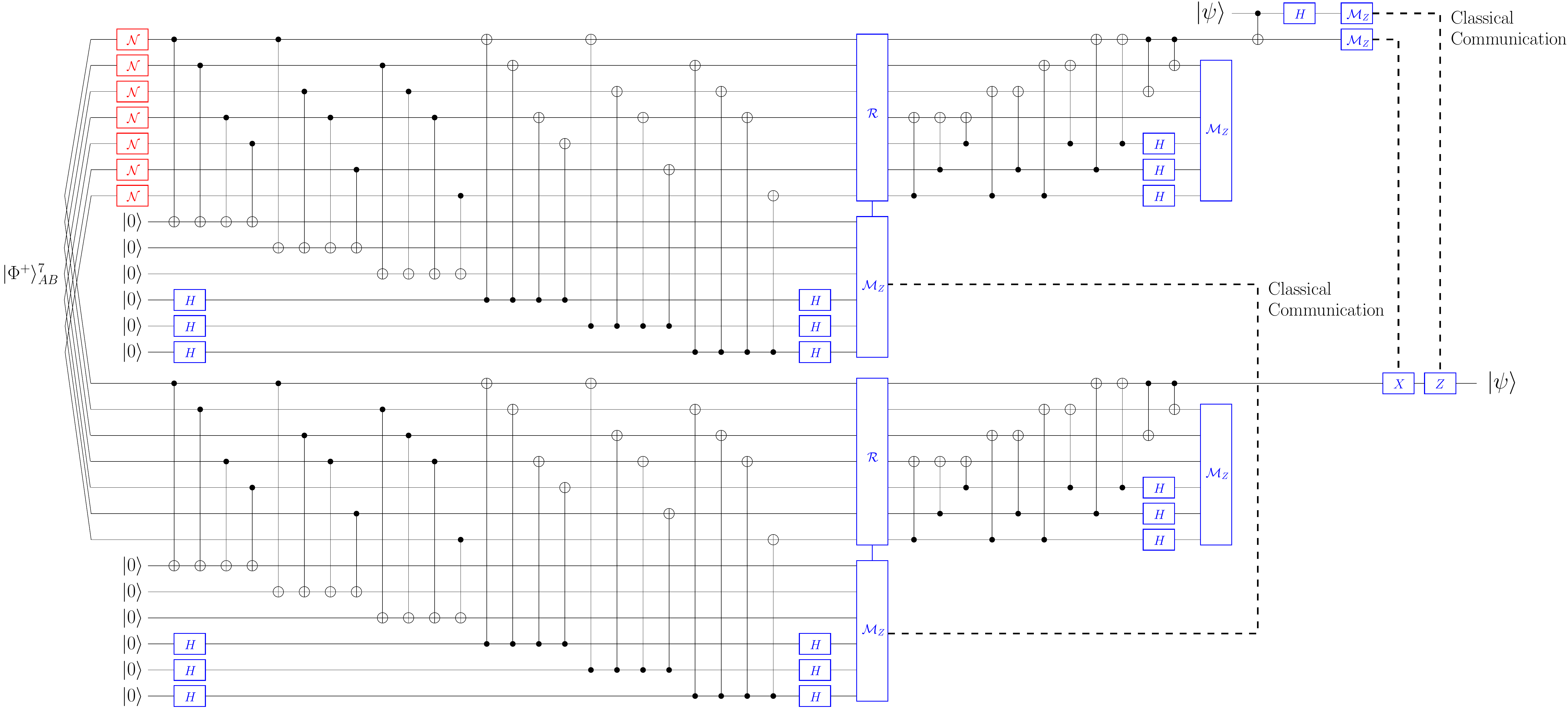}
	\caption{The quantum circuit designed for performing QSC-based QED+QT scheme using the stabilizer operators of Steane code followed by a single qubit quantum teleportation.}
	\label{fig:scheme18}
	\hrulefill
\end{figure*}

\begin{figure*}[t]
	\center
	\includegraphics[width=\linewidth]{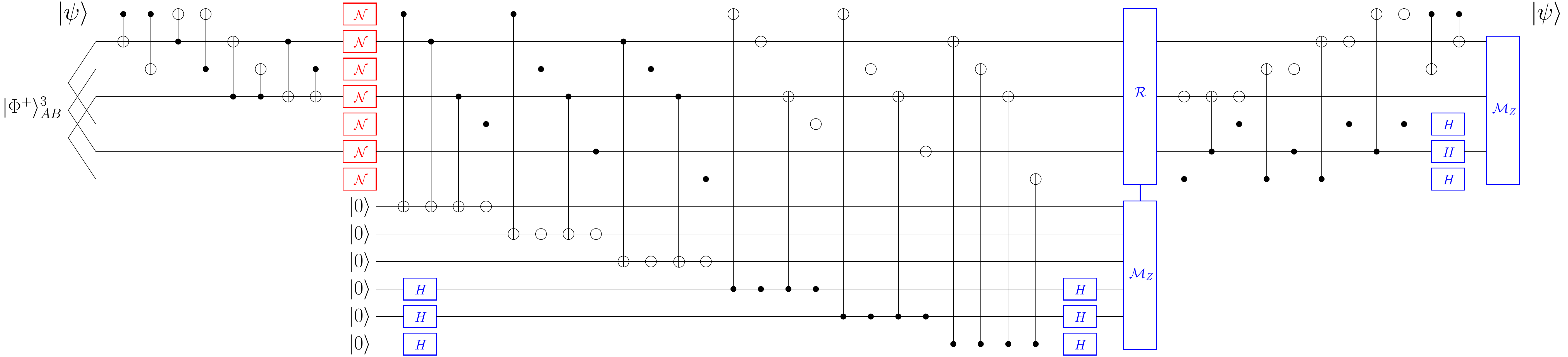}
	\caption{The quantum circuit designed for performing QSC-IE scheme using the stabilizer operators of Steane code.}
	\label{fig:scheme21}
	\hrulefill
\end{figure*}

As for their quantum circuit implementations, our proposed scheme requires a total of 22 CNOT gates as seen in Fig~\ref{fig:encoder}. By contrast, the QSC-based QED+QT scheme requires a total of 71 CNOT gates, namely 48 for stabilizer measurements, 22 for the quantum inverse encoder, and one for quantum teleportation. To elaborate a little further on the quantum circuit implementation required for performing QED+QT scheme using the stabilizer operators of Steane code, please refer to Fig.~\ref{fig:scheme18}. Meanwhile, the QSC-IE scheme requires a total of 43 CNOT gates, namely eight for quantum encoder, 24 for the stabilizer measurements, and 11 for quantum inverse encoder. To provide a clear picture about the quantum circuit implementation of the QSC-IE scheme using the stabilizer operators of Steane code, please refer to Fig.~\ref{fig:scheme21}. We summarize all the physical resources required for performing the error correction schemes to achieve reliable quantum communication in the presence of noisy pre-shared EPR pairs in Table~\ref{table:result2}.

\begin{table*}
	\caption{Comparison of our proposed error-correction scheme with the existing approaches.}
	\small
	\centering
	\begin{tabular}{|c|r|r|r|r|r|}
		\hline
		\multirow{2}{*}{Scheme} & Quantum channel use & Logical qubits & EPR pairs & Classical channel use & \multirow{2}{*}{CNOT} \\
 		& \multicolumn{1}{c|}{$n$} & \multicolumn{1}{c|}{$k$} & \multicolumn{1}{c|}{$e$} & \multicolumn{1}{c|}{$c$} & \\
		\hline
		\hline
		QED+QT & 7 & 1 & 7 & 8 & 71 \\
		\hline
		QSC-IE & 7 & 1 & 3 & 0 & 43 \\
		\hline
		Proposed & 7 & 1 & 6 & 6 & 22 \\
		\hline
	\end{tabular}
	\label{table:result2}
\end{table*}

\begin{remark}
	While attaining an identical QBER and yield to that of QSC-based QED+QT scheme, our proposed arrangement requires fewer pre-shared EPR pairs, fewer classical channel uses, and fewer CNOT gates. However, the proposed arrangement requires more pre-shared EPR pairs and more classical channel uses than the QSC-IE scheme in exchange for fewer CNOT gates and the same number of quantum channel uses.
\end{remark}


\section{Discussion: A Quantum Computing Perspective}
\label{Discussion: A Quantum Computing Perspective}

In the previous sections, we have shown the advantages of our proposal in \textit{quantum communication} applications. In this section, we demonstrate that the proposed scheme can also be adopted for \textit{quantum computing} applications. In quantum computing applications, the quantum information is usually protected with the aid of noise-free auxiliary qubits, which may also take form of pre-shared entanglement~\cite{brun2006correcting, devetak2009entanglement, fujiwara2010entanglement, wilde2012quantum, wilde2013entanglement, grassl2016entanglement}. A prime example is constituted by the family of entanglement-assisted quantum stabilizer codes (EA-QSCs). Compared to the conventional QSCs, which are unassisted by noise-free pre-shared entanglement, EA-QSCs offer an error-correction capability improvement. This is reminiscent of having an additional error-free side channel between the transmitter and the receiver in the classical domain. The argument that we can always have noise-free pre-shared entanglement relies on the assumption that EPR pairs can be created abundantly and quantum entanglement distillation can be applied to them. The concept of EA-QSCs is favourable in the realms of quantum computation, since the EA-QSCs can be readily amalgamated both with transversal implementation of quantum gates~\cite{preskill1998reliable, campbell2017roads} as well as with magic state distillation~\cite{bravyi2005universal} for creating a universal set of fault-tolerant quantum gates. In the following, we propose an error-correction scheme that outperforms the state-of-the-art EA-QSC. 

Any EA-QSC can be defined as $\mathcal{C}[n,k,d,e]$, where $n$ is the number of physical qubits, $k$ is the number of logical qubits, $d$ is the minimum distance of the code, and $e$ is the number of noise-free pre-shared maximally-entangled qubits. The error-detection and error-correction capability of any EA-QSC can be determined by its minimum distance $d$. An EA-QSC exhibiting a minimum distance $d$ is capable of detecting $(d-1)$ quantum errors or correcting $t = \lfloor (d-1)/2 \rfloor$ quantum errors. Based on the quantum Singleton bound of EA-QSCs~\cite{brun2006correcting}, there exists a EA-QSC capable of correcting a single-qubit error $(d = 3)$, which encodes one logical qubit $(k = 1)$ into three physical qubits $(n = 3)$ with the aid of two noise-free pre-shared maximally-entangled qubits $(e = 2)$. This specific code is denoted by $\mathcal{C}[n,k,d,e] = \mathcal{C}=[3,1,3,2]$. In the following, we will show that by utilizing two noise-free pre-shared EPR pairs, instead of error-correction, we can achieve error elimination, implying that in this specific context, we can always obtain a noise-free logical qubit.

\begin{figure*}[t]
    \centering
    \begin{subfigure}{0.49\linewidth}
        \includegraphics[width=\linewidth]{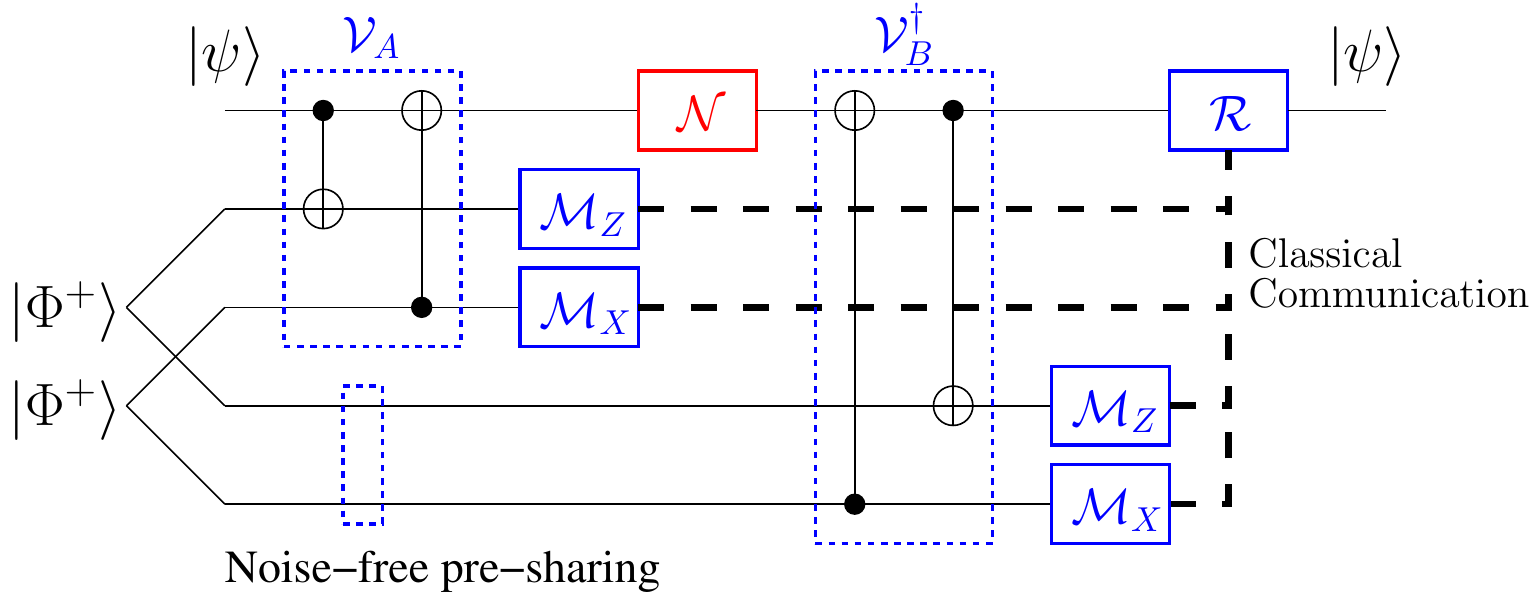}
		\caption{}
		\label{fig:scheme01_ea}
    \end{subfigure}
    \hfill
    \begin{subfigure}{0.49\linewidth}
        \includegraphics[width=\linewidth]{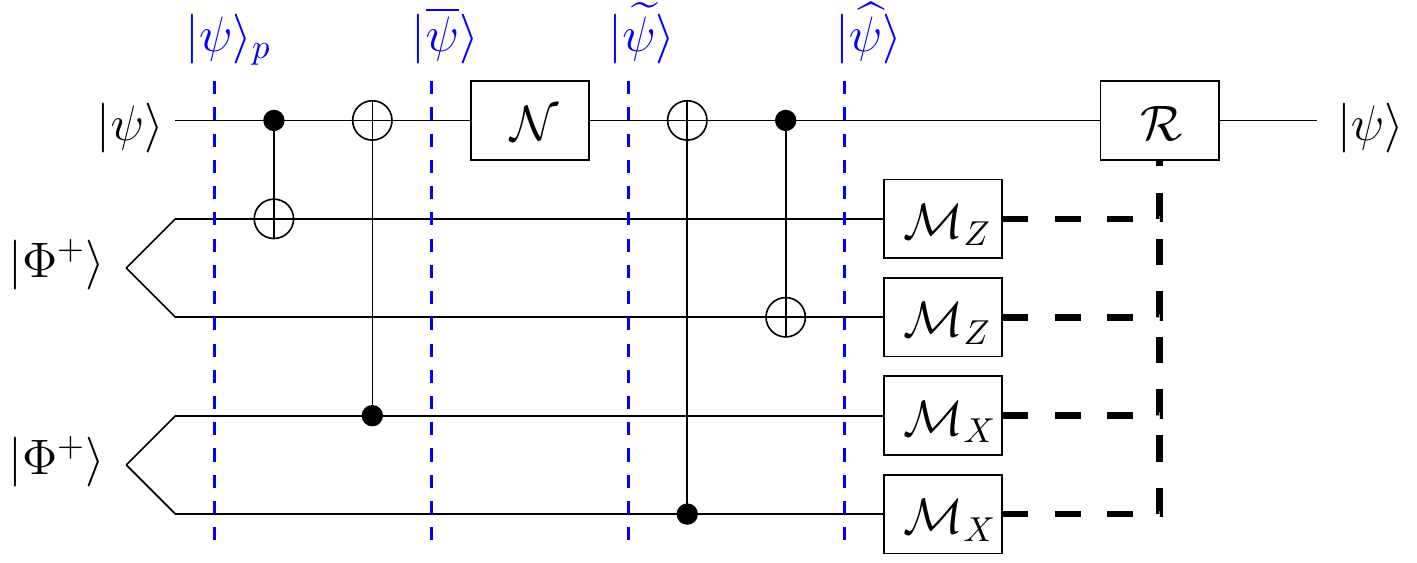}
		\caption{}
		\label{fig:scheme03_ea}
    \end{subfigure}
\caption{(a) The quantum circuit of the proposed scheme utilizing two noise-free pre-shared EPR pairs. (b) The rearranged quantum circuit of (a) for analysis.}
\label{fig:scheme0103_ea}
\hrulefill
\end{figure*}

Let us now discuss our proposed scheme portrayed in Fig.~\ref{fig:scheme0103_ea}(a), which is rearranged into Fig.~\ref{fig:scheme0103_ea}(b) for facilitating our analysis. The quantum channel $\mathcal{N}(\cdot)$ in Fig.~\ref{fig:scheme0103_ea}(a) and~\ref{fig:scheme0103_ea}(b) represents a quantum channel contaminating the logical qubit. According to Fig.~\ref{fig:scheme0103_ea}(b), the quantum encoder $\mathcal{V}_A$ is represented by the following unitary matrix:
\begin{align}
	V_A &= \left( \ket{0} \bra{0} \otimes I \otimes I \otimes I \otimes I + \ket{1} \bra{1} \otimes X \otimes I \otimes I \otimes I \right) \nonumber \\
	& \left( I \otimes I \otimes I \otimes \ket{0} \bra{0} \otimes I + X \otimes I \otimes I \otimes \ket{1} \bra{1} \otimes  I \right),
	\label{eq:va1}
\end{align}
while the quantum decoder $\mathcal{V}_B^{\dagger}$ is described by the following unitary matrix:
\begin{align}
	V_B &= \left( I \otimes I \otimes I \otimes I \otimes \ket{0} \bra{0} + X \otimes I \otimes I \otimes I \otimes \ket{1} \bra{1} \right) \nonumber \\
	& \left( I \otimes I \otimes I \otimes \ket{0} \bra{0} \otimes I + X \otimes I \otimes I \otimes \ket{1} \bra{1} \otimes  I \right).
	\label{eq:vb1}
\end{align}
It can be readily verified that the reversible property is satisfied, i.e. we have $\mathcal{V}_B^{\dagger} \mathcal{V}_A \left( \ket{\psi} \otimes \ket{\Phi^+}_{AB}^2 \right) = \ket{\psi} \otimes \ket{\Phi^+}_{AB}^2 $.

Upon denoting the density matrix of the initial global quantum state of $\ket{\psi} \otimes \ket{\Phi^+}_{AB}^2$ by $\overline{\rho}$, the proposed scheme can be formulated with the aid of the following supermap:
\begin{equation}
	\mathcal{S}(\mathcal{V}_A,\mathcal{N}, \mathcal{V}_B^{\dagger},\overline{\rho}) = \sum_i (V_B N_i V_A)\overline{\rho} (V_B N_i V_A)^{\dagger},
	\label{eq:supermap2}
\end{equation}
where $N_i$ is the Kraus operator describing the quantum channel, while $V_A$ and $V_B$ represent the unitary matrices of~\eqref{eq:va1} and~\eqref{eq:vb1}. Therefore,~\eqref{eq:supermap2} can be rewritten as:
\begin{align}
	\mathcal{S}(\overline{\rho}) &= (1-p)\rho \otimes \ket{\Phi^+} \bra{\Phi^+} \otimes \ket{\Phi^+} \bra{\Phi^+} \nonumber \\
	&+ \frac{p}{3} \left( X\rho X\right) \otimes \ket{\Psi^+} \bra{\Psi^+} \otimes \ket{\Phi^+} \bra{\Phi^+} \nonumber \\ 
 	&+ \frac{p}{3} \left( Y\rho Y\right) \otimes \ket{\Psi^+} \bra{\Psi^+} \otimes \ket{\Phi^-} \bra{\Phi^-} \nonumber \\
 	&+ \frac{p}{3} \left( Z\rho Z\right) \otimes \ket{\Phi^+} \bra{\Phi^+} \otimes \ket{\Phi^-} \bra{\Phi^-}. 
	\label{eq:error-free}
\end{align}

After the decoding operation, we perform the measurement of the EPR pairs. Observe that we can apply $Z$ basis measurement to the first EPR pair and $X$ basis measurement to the second EPR pair for determining the type of Pauli error experienced by the logical qubit $\ket{\psi}$. To elaborate a little further, we design a scheme so that requiring a joint measurement of the EPR pairs can be avoided to reduce the complexity of the quantum encoder and decoder. We combine the classical bits of $A$ and $B$ of Fig.~\ref{fig:scheme0103_ea}(b) to determine the error recovery operator $\mathcal{R}$.

\begin{table}[b]
	\caption{Syndrome values and associated error recovery $\mathcal{R}$ for the scheme in Fig.~\ref{fig:scheme0103_ea}(a).\label{table:syndrome_noisefree}}
	\centering
	\begin{tabular}{|c|c|c|}
		\hline
		$s_Z$ & $s_X$ & Error recovery $\mathcal{R}$ \\
		\hline
		0 & 0 & $I$ \\
		\hline
		1 & 0 & $X$ \\
		\hline
		1 & 1 & $Y$ \\
		\hline
		0 & 1 & $Z$ \\
		\hline
	\end{tabular}
\end{table}

To expound a little further, let us denote the syndrome string as $\underline{s} = s_Z s_X$, where $s_Z$ is obtained from the measurement of the first EPR pair in $Z$ basis and $s_X$ is gleaned from the measurement of the second EPR pair in $X$ basis. The error recovery operator associated with the syndrome value $\underline{s} = s_Z s_X$ is portrayed in Table~\ref{table:syndrome_noisefree}. Finally, it may be inferred from~\eqref{eq:error-free}, that after the error recovery operator $\mathcal{R}$ of Fig.~\ref{fig:scheme0103_ea}(a), we always obtain the legitimate quantum state $\rho$ of the logical qubit. Hence, we have demonstrated that with the aid of two noise-free EPR pairs, instead of correcting a single-qubit achievable by an EA-QSC, we can always recover a noise-free logical qubit. Observe that when we replace the quantum channel $\mathcal{N}(\cdot)$ by realistic noisy quantum Pauli gates, we can modify the LUT of Table~\ref{table:syndrome_noisefree} to benefit from the noise-free operation of the quantum Pauli gates. 



\section{Conclusions and Future Research}
\label{Conclusions and Future Works}

In this treatise, we have conceived a novel direct quantum communication scheme using noisy pre-shared EPR pairs. Conventionally, achieving a reliable quantum communication tends to rely on the consecutive steps of QED followed by quantum teleportation (QED+QT). One of the salient benefits that we can offer is the elimination of the long communication delay imposed by the aforementioned consecutive steps, despite relying on noisy pre-shared EPR pairs. Additionally, our proposed schemes offer better QBER than the recurrence-based QED+QT schemes and provide identical QBER and yield to the QSC-based QED+QT schemes. Moreover, compared to the QSC-based QED+QT schemes, our proposal requires fewer pre-shared entanglement, fewer classical channel uses, and fewer CNOT gates. We have also included the quantum stabilizer code using imperfect pre-shared entanglement (QSC-IE) scheme as our benchmark. Our results show that despite attaining the same level of error-detection and error-correction capability, our proposed scheme requires more pre-shared EPR pairs and more classical channel uses, however, in exchange for fewer CNOT gates requirement and the same number of quantum channel uses. Finally, we have also compared our proposed scheme to EA-QSC, which requires noise-free pre-shared EPR pairs. Again, EA-QSCs require joint eigenvalue measurements relying on all the qubits gleaned from the EPR pairs for performing error-correction. Despite relying only on the local measurements of the EPR pairs and classical communications, we can always obtain a noise-free logical qubit using our proposed scheme.

In our future research, we are interested in finding a systematic way of constructing the quantum encoder and decoder pair. In fact, we found that an arbitrary quantum encoder and decoder pair cannot always satisfy the reversible property of~\eqref{eq:reversible}. Therefore, the sufficient and necessary conditions of generating the quantum encoder and decoder pair should be found. Since our proposed scheme performs identically to the QSC-based QED+QT schemes, it remains to be shown whether a wider range of QSCs can be directly embedded into our scheme. Furthermore, since our proposed scheme requires fewer CNOT gates compared to all state-of-the-art schemes, we are also interested in investigating the performance of the quantum communication schemes under additional realistic assumption of having imperfect quantum gates as well as imperfect measurements and looking at the possibility of creating a fault-tolerant quantum communication protocol.


\section{Appendix A: Proof of Proposition 1}
\label{Proof of Proposition 1}

By exploiting the quantum depolarizing channel model of Section~\ref{System Model} and by utilizing the expressions of~\eqref{eq:va},~\eqref{eq:supermap1} can be reformulated as shown in~\eqref{eq:result1}, where $\rho$ is the density matrix of the logical qubit and we assume that the quantum depolarizing channels experienced by $\ket{\psi}$ and $\ket{\Phi^+}_{A}$ exhibit an identical depolarizing probability $p$.

\begin{figure*}[t]
\setcounter{equation}{28}
\begin{align}
	\mathcal{S}(\overline{\rho}) &= \Big[ (1-p)^2 \rho + \frac{p^2}{9} X\rho X + \frac{p^2}{9} Y\rho Y + \frac{p(1-p)}{3} Z\rho Z \Big] \otimes \ket{\Phi^+} \bra{\Phi^+} \nonumber \\
 	&+ \Big[ \frac{p^2}{9} \rho + \frac{p^2}{9} X\rho X + \frac{p^2}{9} Y\rho Y + \frac{p(1-p)}{3} Z\rho Z \Big] \otimes \ket{\Phi^-} \bra{\Phi^-} \nonumber \\
	&+ \Big[ \frac{p^2}{9} \rho + \frac{p^2}{9} X\rho X + \frac{p^2}{9} Y\rho Y + \frac{p(1-p)}{3} Z\rho Z \Big]\otimes \ket{\Psi^-} \bra{\Psi^-} \nonumber \\
 	&+ \Big[ \frac{p(1-p)}{3} \rho + \frac{p(1-p)}{3} X\rho X + \frac{p(1-p)}{3} Y\rho Y + \frac{p^2}{9} Z\rho Z \Big] \otimes \ket{\Psi^+} \bra{\Psi^+},
\label{eq:result1}
\end{align}
\hrulefill
\end{figure*}

After the decoding, a measurement in the $Z$ basis of the EPR pair shared between $A$ and $B$ is performed. Every time we find a disagreement in the classical measurement results from the EPR pair $(s = s_A \oplus s_B = 1)$, the associated logical qubit is discarded, otherwise, it is retained. We note that the syndrome value of $s = 0$ is obtained if the EPR pair is in the quantum state $\ket{\Phi^+}$ or $\ket{\Phi^-}$, while the EPR pair in the state $\ket{\Psi^+}$ or $\ket{\Psi^-}$ gives us a syndrome value of $s = 1$. Hence, the probability of retaining the logical qubit is equal to the probability of obtaining the syndrome value $s = 0$. Based on these considerations and by accounting for~\eqref{eq:result1}, we can determine the probability of obtaining the syndrome value $s = 0$:
\begin{equation}
	p(s = 0) = 1 - \frac{4p}{3} + \frac{8p^2}{9},
\end{equation}
which is obtained from the following error operators $P \in \lbrace II, IZ, XX, XY, YX, YY, ZI, ZZ \rbrace$. Then, based on this set of error operators, we can determine the probability of obtaining the syndrome value of $s = 0$ and obtain the legitimate quantum state of the logical qubit $\rho$ : 
\begin{equation}
	p(\rho \cap (s = 0)) = 1 -2p + \frac{10p^2}{9},
\end{equation}
which is obtained from the following error operators $P \in \lbrace II, ZZ \rbrace$. More specifically, the error operator $ZZ$ imposed by the quantum channels is transformed into an error operator $IZ$ at the output of the quantum decoder $\mathcal{V}^{\dagger}_B$ of Fig.~\ref{fig:scheme15}. Consequently, we have the final quantum state of $\rho \otimes \ket{\Phi^-} \bra{\Phi^-}$. Since the measurement of the EPR pair is performed in $Z$ basis, we obtain the syndrome value of $s = 0$, while retaining the legitimate quantum state of the logical qubit $\rho$. Finally, the success probability of the scheme presented in Fig.~\ref{fig:scheme15} can be determined according to Definition~\ref{def:1} as follows:
\begin{align}
	p_s &= p(\rho | (s = 0)) = \frac{1-2p+\frac{10p^2}{9}}{1-\frac{4p}{3}+\frac{8p^2}{9}} \nonumber \\
	&= 1 - \frac{2p}{3} - \frac{2p^2}{3} - \frac{8p^3}{27} + \frac{16p^4}{81} + \mathcal{O}(p^5),
\end{align}
which gives us an approximately linear performance improvement over the uncoded QBER as a function of $p$. By accounting for Definition~\ref{def:2}, the yield is $Y = p(s = 0)$ and the proof follows.

\section{Appendix B: Proof of Proposition 2}
\label{Proof of Proposition 2}

After applying the first CNOT of the decoder $\mathcal{V}_B^{\dagger}$ of Fig.~\ref{fig:scheme16}, we can determine the probability of obtaining the syndrome value $s_Z = 0$ from the first EPR pair as follows:
\begin{equation}
	p(s_Z = 0) = 1 -\frac{4p}{3} +\frac{8p^2}{9}.
\end{equation}
If $s_Z = 1$, we discard the logical qubit. If $s_Z = 0$, the first-two quantum depolarizing channels are reduced into a single depolarizing channel having the following Kraus operators: 
\begin{align}
	& N_1 = \sqrt{\frac{1-2p+\frac{10p^2}{9}}{1-\frac{4p}{3}+\frac{8p^2}{9}}}I, & N_2 = \sqrt{\frac{\frac{2p^2}{9}}{1-\frac{4p}{3}+\frac{8p^2}{9}}}X, \nonumber \\  
	& N_3 = \sqrt{\frac{\frac{2p^2}{9}}{1-\frac{4p}{3}+\frac{8p^2}{9}}}Y, & N_4 = \sqrt{\frac{\frac{2p}{3}-\frac{2p^2}{3}}{1-\frac{4p}{3}+\frac{8p^2}{9}}}Z.
\end{align}
Now, by applying the second CNOT of the decoder $\mathcal{V}_B^{\dagger}$ of Fig.~\ref{fig:scheme16}, we can determine the probability of obtaining the syndrome value $s_X = 0$ from the second EPR pair as follows:
\begin{equation}
	p(s_X = 0) = \frac{1 - \frac{8p}{3} + \frac{28p^2}{9} - \frac{32p^3}{27}} {1-\frac{4p}{3}+\frac{8p^2}{9}}.
\end{equation}
If $s_Z = 1$, we discard the logical qubit, while If $s_Z = 0$, we retain the logical qubit. Therefore, the probability of arriving at the legitimate quantum state $\rho$ and measuring the syndrome value $s_X = 0$ is given by 
\begin{equation}
	p(\rho \cap (s_X = 0)) = \frac{1 - 3p + \frac{28p^2}{9} - \frac{28p^3}{27}}{1-\frac{4p}{3}+\frac{8p^2}{9}}.	
\end{equation}
Finally, by using Definition~\ref{def:1} and~\ref{def:2}, the proof follows.

\section{Appendix C: Proof of Proposition 3}
\label{Proof of Proposition 3}

Similar to the proof of Proposition~\ref{prop_1}, By exploiting the error model of Section~\ref{System Model} and relying on the quantum encoder and decoder of Fig.~\ref{fig:scheme13}, the supermap of~\eqref{eq:supermap1} can be readily obtained. After the decoding operation, the first EPR pair is measured in the $X$ basis while the second one in the $Z$ basis. Let us distinguish the components of the syndrome string in~\eqref{eq:sindrome} according to the basis used for the measurement. Specifically, let us denote the syndrome component obtained when the second EPR pair is measured in the $Z$ basis by $s_Z = s_A \oplus s_B$, while the syndrome component obtained when the first EPR pair is measured in the $X$ basis by $s_X = s_A \oplus s_B$. The overall syndrome string is $\underline{s} = s_X s_Z$. Since no error operators exhibiting even numbers of $X$ errors and even numbers of $Z$ errors can be detected, which gives us the syndrome vector of $\underline{s} =  00$, the probability we retain the logical qubits is equal to the sum of the probabilities of all these possible error patterns resulting in the syndrome vector of $\underline{s} = 00$. After observing $4^4 = 256$ error patterns, we found 64 error patterns that generate the syndrome vector of $\underline{s} = 00$: one all-identity error pattern (weight = 0); 18 error patterns having Pauli weight = 2; 24 error patterns having Pauli weight = 3; and 21 error patterns having Pauli weight = 4. Therefore, by accounting for~\eqref{eq:supermap1}, we have 
\begin{align}
	p(\underline{s} = 00) &= \sum_{i = 0}^{n} W_i \left( \frac{p}{3} \right)^i (1-p)^{(n -i)} \nonumber \\ 
	&= 1 - 4p + 8p^2 - \frac{64p^3}{9} + \frac{64p^4}{27},
\end{align}
where $W = \lbrace W_0, W_1, W_2, W_3, W_4 \rbrace = \lbrace 1, 0, 18, 24, 21 \rbrace$. However, from all of those error patterns, only four are actually associated with the legitimate quantum state $\rho$ of the logical qubits, which are $P \in \lbrace IIII, XXXX, YYYY, ZZZZ \rbrace$. Hence, the probability of obtaining the legitimate quantum state $\rho$ while measuring $\underline{s} = 00$ is given by 
\begin{align}
	p(\rho \cap (\underline{s} = 00)) &= (1-p)^4 + 3\left(\frac{p}{3}\right)^4 \nonumber \\
	&= 1 -4p + 6p^2 - 4p^3 + \frac{28p^4}{27}.
\end{align}
Finally, by using Definition~\ref{def:1} and~\ref{def:2}, the proof follows.

\bibliographystyle{ieeetr}

\begin{thebibliography}{10}

\bibitem{kimble2008quantum}
H.~J. Kimble, ``{The Quantum Internet},'' {\em Nature}, vol.~453, no.~7198,
  pp.~1023--1030, 2008.

\bibitem{caleffi2018quantum}
M.~Caleffi, A.~S. Cacciapuoti, and G.~Bianchi, ``{Quantum Internet: From
  communication to distributed computing!},'' in {\em Proceedings of the 5th
  ACM International Conference on Nanoscale Computing and Communication},
  pp.~1--4, 2018.

\bibitem{wehner2018quantum}
S.~Wehner, D.~Elkouss, and R.~Hanson, ``{Quantum Internet: A vision for the
  road ahead},'' {\em Science}, vol.~362, no.~6412, 2018.

\bibitem{qirg-use}
C.~Wang, A.~Rahman, and R.~Li, ``{Applications and use cases for the Quantum
  Internet},'' Internet Draft draft-wang-qirg-quantum-internet-use-cases-04,
  Internet Engineering Task Force, Mar. 2020.
\newblock Work in Progress.

\bibitem{razavi2012multiple}
M.~Razavi, ``{Multiple-access quantum key distribution networks},'' {\em IEEE
  Transactions on Communications}, vol.~60, no.~10, pp.~3071--3079, 2012.

\bibitem{cuomo2020towards}
D.~Cuomo, M.~Caleffi, and A.~S. Cacciapuoti, ``{Towards a distributed quantum
  computing ecosystem},'' {\em IET Quantum Communication (Invited Paper)},
  vol.~1, no.~1, pp.~3--8, 2020.

\bibitem{caleffi2020rise}
M.~Caleffi, D.~Chandra, D.~Cuomo, S.~Hassanpour, and A.~S. Cacciapuoti, ``{The
  rise of the Quantum Internet},'' {\em Computer}, vol.~53, no.~6, pp.~67--72,
  2020.

\bibitem{sun2020toward}
Z.~Sun, L.~Song, Q.~Huang, L.~Yin, G.~Long, J.~Lu, and L.~Hanzo, ``{Toward
  practical quantum secure direct communication: A quantum-memory-free protocol
  and code design},'' {\em IEEE Transactions on Communications}, vol.~68,
  no.~9, pp.~5778--5792, 2020.

\bibitem{cacciapuoti2019quantum}
A.~S. Cacciapuoti, M.~Caleffi, F.~Tafuri, F.~S. Cataliotti, S.~Gherardini, and
  G.~Bianchi, ``{Quantum Internet: Networking challenges in distributed quantum
  computing},'' {\em IEEE Network}, vol.~34, no.~1, pp.~137--143, 2019.

\bibitem{cacciapuoti2020entanglement}
A.~S. Cacciapuoti, M.~Caleffi, R.~Van~Meter, and L.~Hanzo, ``{When entanglement
  meets classical communications: Quantum teleportation for the Quantum
  Internet},'' {\em IEEE Transactions on Communications (Invited Paper)},
  vol.~68, no.~6, pp.~3808--3833, 2020.

\bibitem{nielsen2000quantum}
M.~A. Nielsen and I.~L. Chuang, {\em {Quantum computation and quantum
  information}}.
\newblock Cambridge University Press, 2000.

\bibitem{cariolaro2010performance}
G.~Cariolaro and G.~Pierobon, ``{Performance of quantum data transmission
  systems in the presence of thermal noise},'' {\em IEEE Transactions on
  Communications}, vol.~58, no.~2, pp.~623--630, 2010.

\bibitem{shannon1948mathematical}
C.~E. Shannon, ``{A mathematical theory of communication},'' {\em The Bell
  System Technical Journal}, vol.~27, no.~3, pp.~379--423, 1948.

\bibitem{shor1995scheme}
P.~W. Shor, ``{Scheme for reducing decoherence in quantum computer memory},''
  {\em Physical Review A}, vol.~52, no.~4, 1995.

\bibitem{calderbank1996good}
A.~R. Calderbank and P.~W. Shor, ``{Good quantum error-correcting codes
  exist},'' {\em Physical Review A}, vol.~54, no.~2, 1996.

\bibitem{laflamme1996perfect}
R.~Laflamme, C.~Miquel, J.~P. Paz, and W.~H. Zurek, ``{Perfect quantum error
  correcting code},'' {\em Physical Review Letters}, vol.~77, no.~1, 1996.

\bibitem{lidar2013quantum}
D.~A. Lidar and T.~A. Brun, {\em {Quantum error correction}}.
\newblock Cambridge University Press, 2013.

\bibitem{babar2018duality}
Z.~Babar, D.~Chandra, H.~V. Nguyen, P.~Botsinis, D.~Alanis, S.~X. Ng, and
  L.~Hanzo, ``{Duality of quantum and classical error correction codes: Design
  principles and examples},'' {\em IEEE Communications Surveys \& Tutorials},
  vol.~21, no.~1, pp.~970--1010, 2018.

\bibitem{fujiwara2010entanglement}
Y.~Fujiwara, D.~Clark, P.~Vandendriessche, M.~De~Boeck, and V.~D. Tonchev,
  ``{Entanglement-assisted quantum low-density parity-check codes},'' {\em
  Physical Review A}, vol.~82, no.~4, 2010.

\bibitem{wilde2012quantum}
M.~M. Wilde and J.~M. Renes, ``{Quantum polar codes for arbitrary channels},''
  in {\em Proceedings of the IEEE International Symposium on Information Theory
  (ISIT)}, pp.~334--338, IEEE, 2012.

\bibitem{wilde2013entanglement}
M.~M. Wilde, M.-H. Hsieh, and Z.~Babar, ``{Entanglement-assisted quantum turbo
  codes},'' {\em IEEE Transactions on Information Theory}, vol.~60, no.~2,
  pp.~1203--1222, 2013.

\bibitem{chandra2017quantum}
D.~Chandra, Z.~Babar, H.~V. Nguyen, D.~Alanis, P.~Botsinis, S.~X. Ng, and
  L.~Hanzo, ``{Quantum coding bounds and a closed-form approximation of the
  minimum distance versus quantum coding rate},'' {\em IEEE Access}, vol.~5,
  pp.~11557--11581, 2017.

\bibitem{brun2006correcting}
T.~A. Brun, I.~Devetak, and M.-H. Hsieh, ``{Correcting quantum errors with
  entanglement},'' {\em Science}, vol.~314, no.~5798, pp.~436--439, 2006.

\bibitem{devetak2009entanglement}
I.~Devetak, T.~A. Brun, and M.-H. Hsieh, ``{Entanglement-assisted quantum
  error-correcting codes},'' in {\em New Trends in Mathematical Physics:
  Selected Contributions of the 15th International Congress on Mathematical
  Physics}, pp.~161--172, Springer, 2009.

\bibitem{grassl2016entanglement}
M.~Grassl, ``{Entanglement-assisted quantum communication beating the quantum
  Singleton bound},'' in {\em Proceedings of the 16th Asian Quantum Information
  Science Conference (AQIS)}, pp.~20--21, 2016.

\bibitem{bennett1996concentrating}
C.~H. Bennett, H.~J. Bernstein, S.~Popescu, and B.~Schumacher, ``{Concentrating
  partial entanglement by local operations},'' {\em Physical Review A},
  vol.~53, no.~4, 1996.

\bibitem{bennett1996purification}
C.~H. Bennett, G.~Brassard, S.~Popescu, B.~Schumacher, J.~A. Smolin, and W.~K.
  Wootters, ``{Purification of noisy entanglement and faithful teleportation
  via noisy channels},'' {\em Physical Review Letters}, vol.~76, no.~5, p.~722,
  1996.

\bibitem{bennett1996mixed}
C.~H. Bennett, D.~P. DiVincenzo, J.~A. Smolin, and W.~K. Wootters,
  ``{Mixed-state entanglement and quantum error correction},'' {\em Physical
  Review A}, vol.~54, no.~5, 1996.

\bibitem{matsumoto2003conversion}
R.~Matsumoto, ``{Conversion of a general quantum stabilizer code to an
  entanglement distillation protocol},'' {\em Journal of Physics A:
  Mathematical and General}, vol.~36, no.~29, 2003.

\bibitem{bennett1993teleporting}
C.~H. Bennett, G.~Brassard, C.~Cr{\'e}peau, R.~Jozsa, A.~Peres, and W.~K.
  Wootters, ``{Teleporting an unknown quantum state via dual classical and
  Einstein-Podolsky-Rosen channels},'' {\em Physical Review Letters}, vol.~70,
  no.~13, 1993.

\bibitem{lai2012entanglement}
C.-Y. Lai and T.~A. Brun, ``{Entanglement-assisted quantum error-correcting
  codes with imperfect ebits},'' {\em Physical Review A}, vol.~86, no.~3, 2012.

\bibitem{chandra2019near}
D.~Chandra, Z.~Babar, S.~X. Ng, and L.~Hanzo, ``{Near-hashing-bound
  multiple-rate quantum turbo short-block codes},'' {\em IEEE Access}, vol.~7,
  pp.~52712--52730, 2019.

\bibitem{nguyen2016exit}
H.~V. Nguyen, Z.~Babar, D.~Alanis, P.~Botsinis, D.~Chandra, S.~X. Ng, and
  L.~Hanzo, ``{EXIT-chart aided quantum code design improves the normalised
  throughput of realistic quantum devices},'' {\em IEEE Access}, vol.~4,
  pp.~10194--10209, 2016.

\bibitem{chandra2019quantum}
D.~Chandra, Z.~Babar, H.~V. Nguyen, D.~Alanis, P.~Botsinis, S.~X. Ng, and
  L.~Hanzo, ``{Quantum topological error correction codes are capable of
  improving the performance of Clifford gates},'' {\em IEEE Access}, vol.~7,
  pp.~121501--121529, 2019.

\bibitem{cane2020mitigation}
R.~Cane, D.~Chandra, S.~X. Ng, and L.~Hanzo, ``{Mitigation of
  decoherence-induced quantum-bit errors and quantum-gate errors using
  Steane’s code},'' {\em IEEE Access}, vol.~8, pp.~83693--83709, 2020.

\bibitem{steane1996multiple}
A.~Steane, ``{Multiple-particle interference and quantum error correction},''
  {\em Proceedings of the Royal Society of London - Series A: Mathematical,
  Physical and Engineering Sciences}, vol.~452, no.~1954, pp.~2551--2577, 1996.

\bibitem{preskill1998reliable}
J.~Preskill, ``{Reliable quantum computers},'' {\em Proceedings of the Royal
  Society of London - Series A: Mathematical, Physical and Engineering
  Sciences}, vol.~454, no.~1969, pp.~385--410, 1998.

\bibitem{campbell2017roads}
E.~T. Campbell, B.~M. Terhal, and C.~Vuillot, ``{Roads towards fault-tolerant
  universal quantum computation},'' {\em Nature}, vol.~549, no.~7671,
  pp.~172--179, 2017.

\bibitem{bravyi2005universal}
S.~Bravyi and A.~Kitaev, ``{Universal quantum computation with ideal Clifford
  gates and noisy ancillas},'' {\em Physical Review A}, vol.~71, no.~2, 2005.

\end{thebibliography}

\begin{IEEEbiography}[{\includegraphics[width=1in,height=1.25in,clip,keepaspectratio]{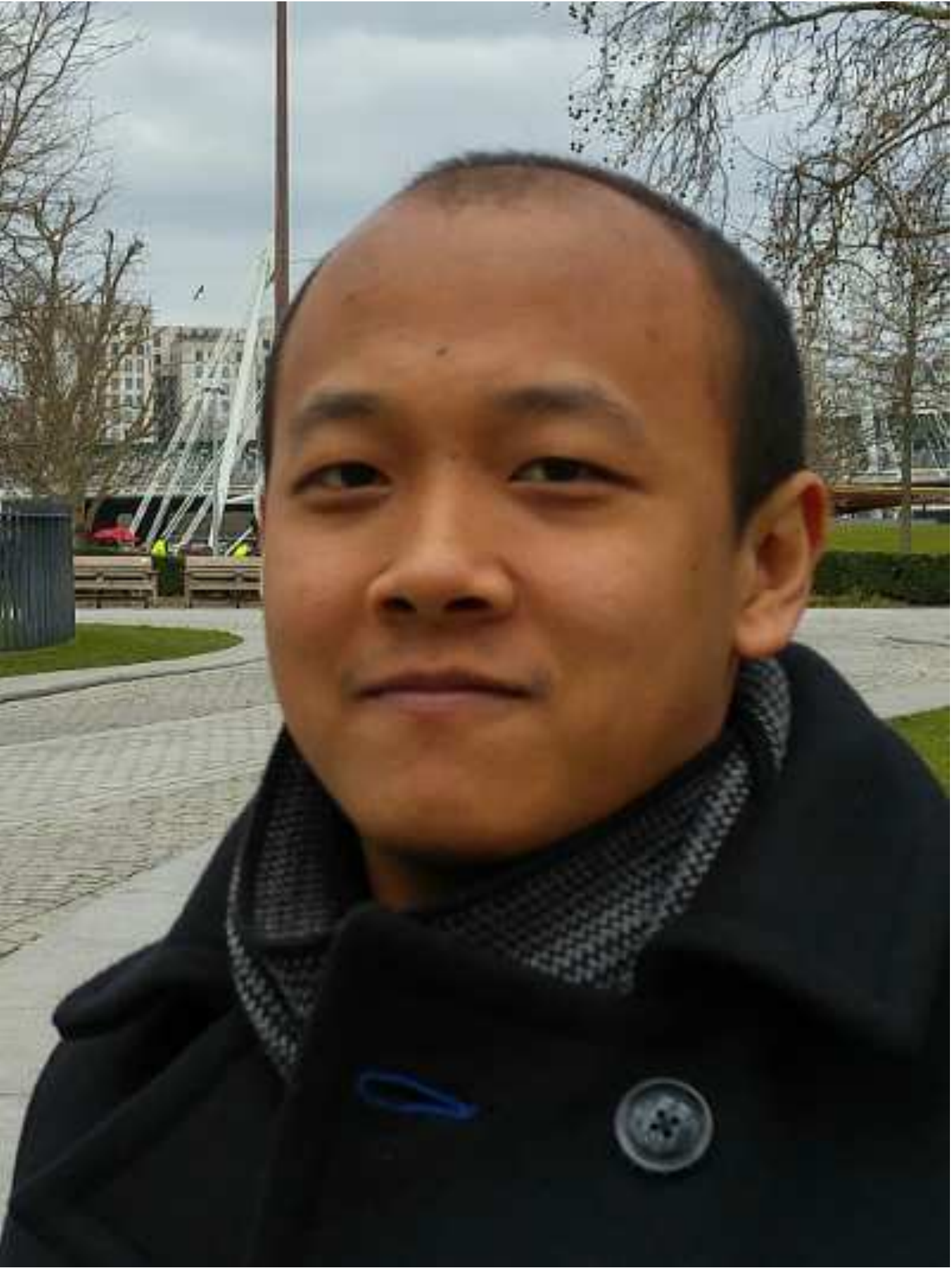}}]{Daryus Chandra} (S'15, M'20) received the M.Eng. degree in electrical engineering from Universitas Gadjah Mada, Indonesia, in 2014 and the Ph.D. degree in electronics and electrical engineering from University of Southampton, UK, in 2020. He was a research fellow with the Future Communications Laboratory, University of Naples Federico II, Italy. Currently, he is a research fellow with the Next-Generation Wireless Research Group, University of Southampton, UK. His current research interests include classical and quantum error-correction codes, quantum information, and quantum communications.
\end{IEEEbiography}

\begin{IEEEbiography}
[{\includegraphics[width=1in,height=1.25in,clip,keepaspectratio]{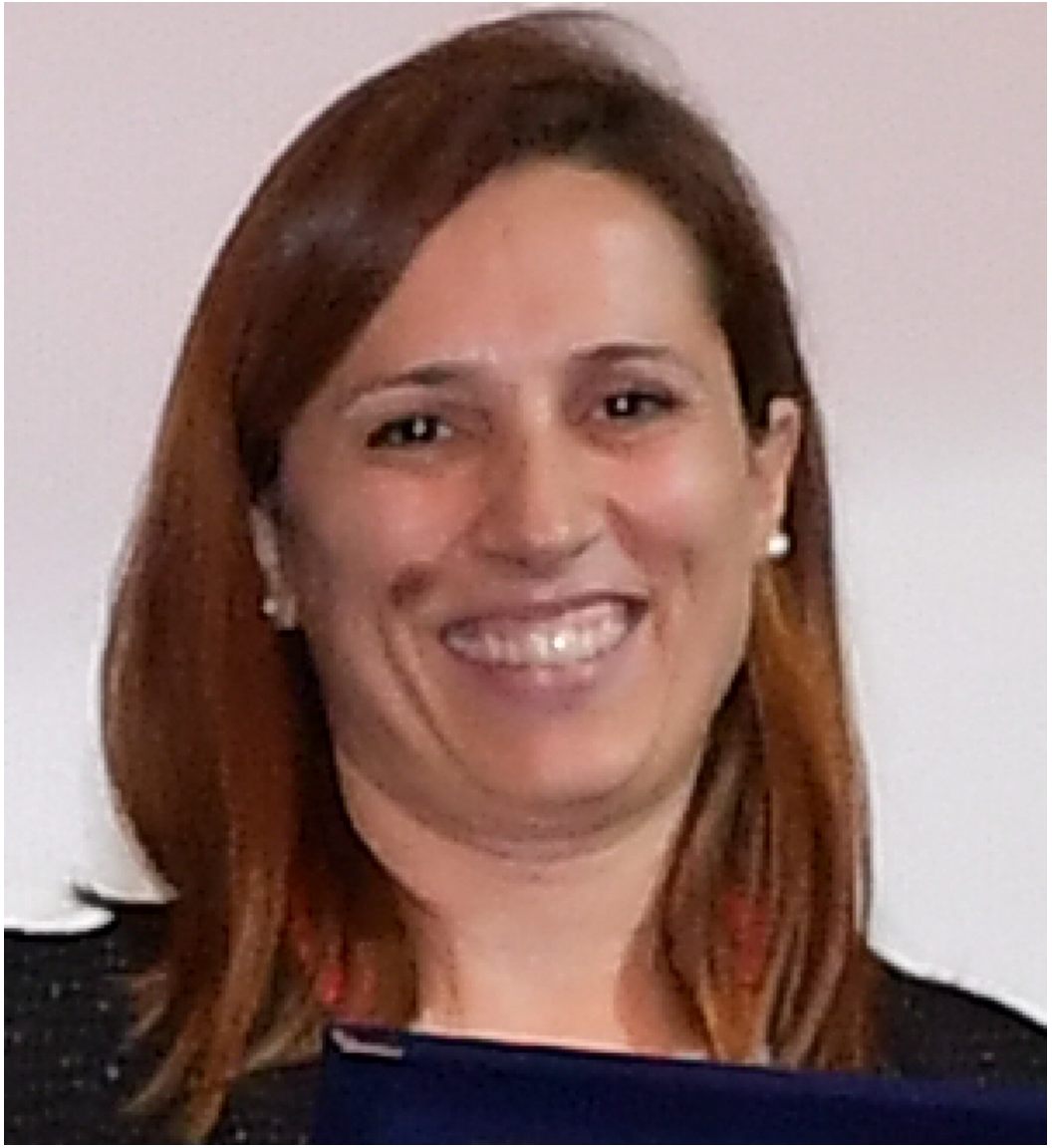}}]{Angela Sara Cacciapuoti} (M'10, SM'16) is an Associate Professor at the University of Naples Federico II, Italy. Since July 2018, she held the national habilitation as ``Full Professor" in Telecommunications Engineering. Her work has appeared in first tier IEEE journals and she has received different awards, including the elevation to the grade of IEEE Senior Member in 2016. Currently, Angela Sara serves as \textit{Area Editor} for IEEE Communications Letters, and as \textit{Editor/Associate Editor} for the journals: IEEE Trans. on Communications, IEEE Trans. on Wireless Communications, IEEE Trans. on Quantum Engineering, IEEE Network and IEEE Open Journal of Communications Society. She was a recipient of the 2017 Exemplary Editor Award of the IEEE Communications Letters. In 2016 she has been an appointed member of the IEEE ComSoc Young Professionals Standing Committee. From 2017 to 2018, she has been the Award Co-Chair of the N2Women Board. From 2017 to 2020, she has been the Treasurer of the IEEE Women in Engineering (WIE) Affinity Group of the IEEE Italy Section. In 2018, she has been appointed as Publicity Chair of the IEEE ComSoc Women in Communications Engineering (WICE) Standing Committee. And since 2020, she is the vice-chair of WICE. Her current research interests are mainly in Quantum Communications, Quantum Networks and Quantum Information Processing.
\end{IEEEbiography}

\begin{IEEEbiography}
[{\includegraphics[width=1in,height=1.25in,clip,keepaspectratio]{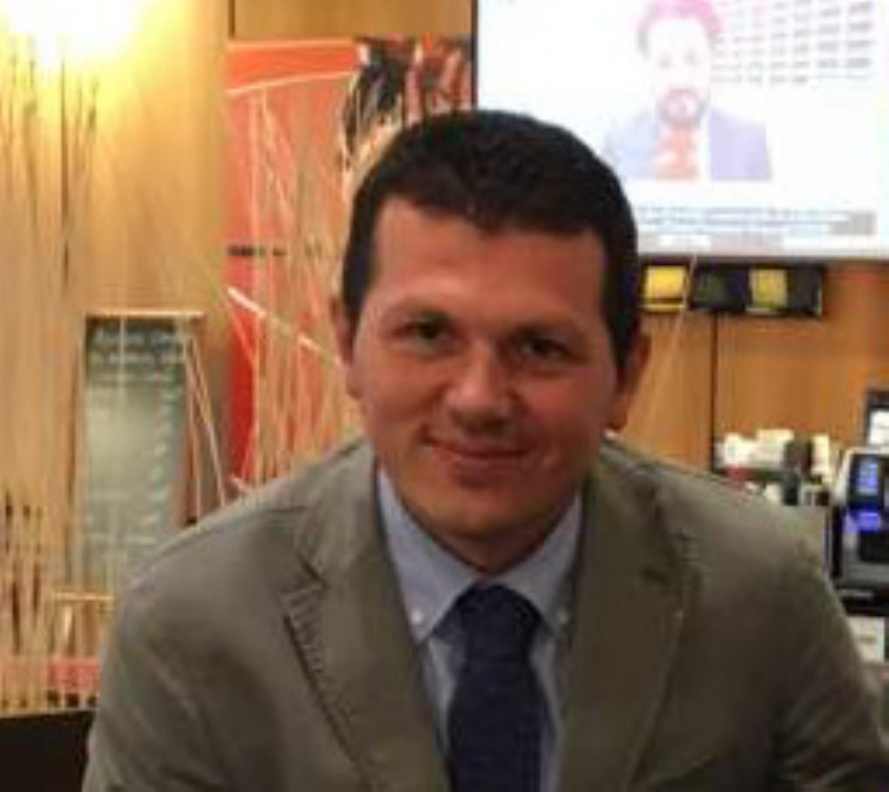}}]{Marcello Caleffi} (M'12, SM'16) is associate professor with DIETI Department, University of Naples Federico II. Previously, he was with the Broadband Wireless Networking Laboratory, Georgia Institute of Technology, Atlanta, GA, USA, and also with the NaNoNetworking Center in Catalunya (N3Cat), Universitat Politecnica de Catalunya (UPC), Barcelona, as visiting researcher. Since July 2018, he held the Italian National Habilitation as \textit{full professor} in telecommunications engineering. He currently serves as \textit{associate editor/associate technical editor} for IEEE Communications Magazine, IEEE Trans. on Wireless Communications, IEEE Trans. on Quantum Engineering and IEEE Communications Letters. In 2017, he has been appointed as a Distinguished Lecturer for the IEEE Computer Society. In December 2018, he has been an appointed member of the IEEE \textit{New Initiatives Committee}.
\end{IEEEbiography}

\begin{IEEEbiography}
[{\includegraphics[width=1in,height=1.25in,clip,keepaspectratio]{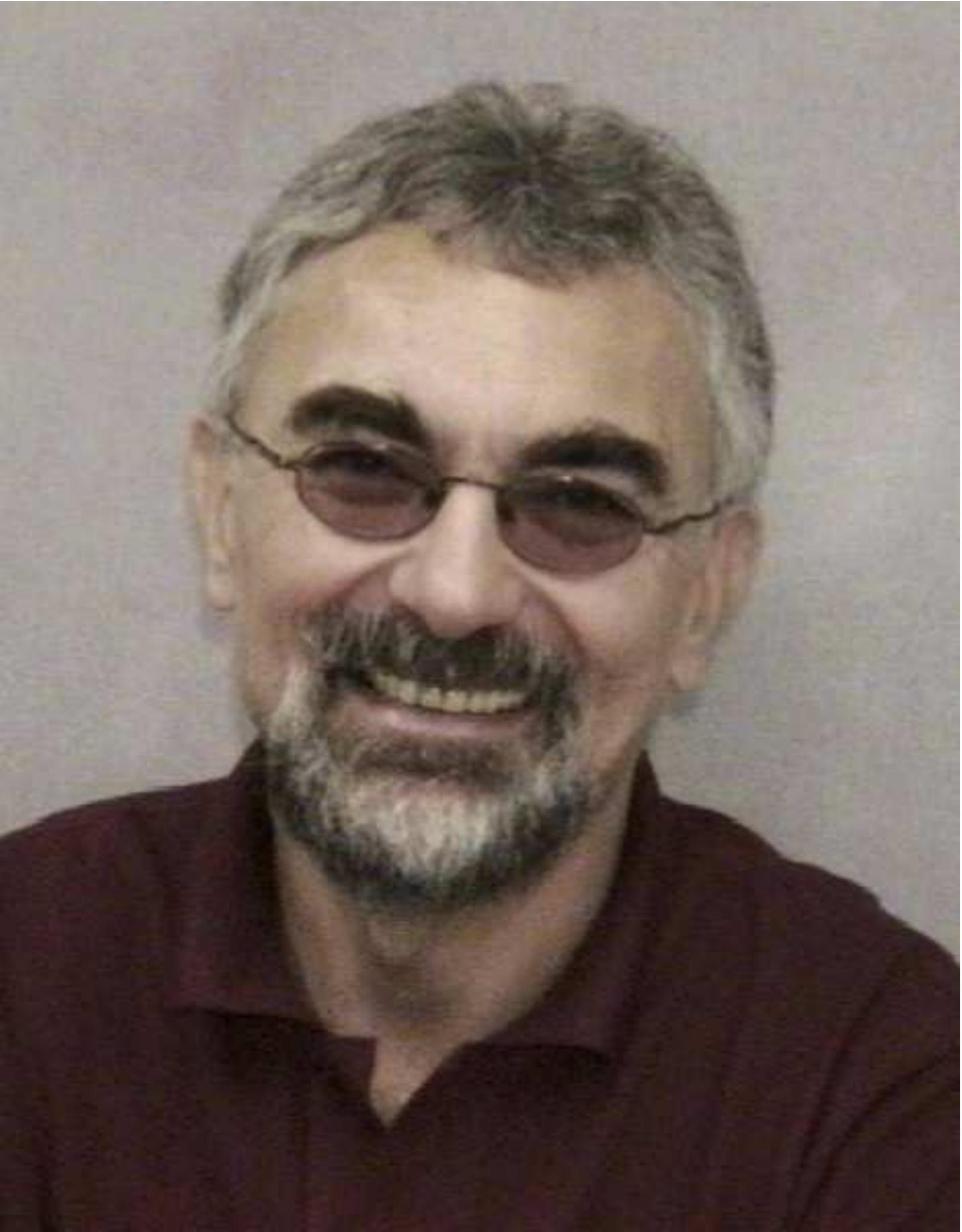}}]{Lajos Hanzo} (FIEEE'04) received his Master degree and Doctorate in 1976 and 1983, respectively from the Technical University (TU) of Budapest. He was also awarded the Doctor of Sciences (DSc) degree
by the University of Southampton (2004) and Honorary Doctorates by the TU of Budapest (2009) and by the University of Edinburgh (2015). He is a Foreign Member of the Hungarian Academy of Sciences and a former Editor-in-Chief of the IEEE Press. He has served several terms as Governor of both IEEE ComSoc and of VTS. He has published 1970 contributions at IEEE Xplore, 19 Wiley-IEEE Press books and has helped the fast-track career of 123 PhD students. Over 40 of them are Professors at various stages of their careers in academia and many of them are leading scientists in the wireless industry. He is also a Fellow of the Royal Academy of Engineering (FREng), of the IET and of EURASIP. (\href{http://www-mobile.ecs.soton.ac.uk}{http://www-mobile.ecs.soton.ac.uk}, \href{https://en.wikipedia.org/wiki/Lajos-Hanzo}{https://en.wikipedia.org/wiki/Lajos-Hanzo})
\end{IEEEbiography}

\end{document}